\documentclass[12pt]{iopart}

\usepackage{graphicx,bbm,amsthm,amsfonts,amssymb,amsopn,amsmath}
\usepackage{dsfont}
\usepackage{iopams}
\usepackage{mathtools}
\usepackage{cite}
\usepackage{hyperref}
\usepackage{pgf,tikz}
\usetikzlibrary{arrows,calc,matrix,backgrounds}

\newcommand{\tikzmark}[2]{\tikz[overlay,remember picture, baseline=(#1.base)]\node(#1){\ensuremath{#2}};}
\newcommand\scalemath[2]{\scalebox{#1}{\mbox{\ensuremath{\displaystyle #2}}}}
\def\be{\begin{equation}}
\def\ee{\end{equation}}

\def\ul#1{{\underline{#1}}}
\newcommand{\un}[1]{{{#1}'}}

\newcommand{\emptysub}[1]{\vphantom{]}_{{\raisebox{-1.75pt}{$\scriptstyle #1$}}}}

\newcommand{\bea}{\begin{eqnarray}}
\newcommand{\eea}{\end{eqnarray}}

\newcommand{\ave}[1]{{\langle #1\rangle}}

\newcommand{\ii}{ {\rm i} }
\newcommand{\dd}{ {\rm d} }

\newcommand{\ZZ}{\mathbb{Z}}

\newcommand{\x}{{\rm x}}

\newcommand{\oo}[1]{{[ #1 ]}}

\newcommand{\lsp}{\text{lsp}}
\newcommand{\odd}{{\mathrm o}}
\newcommand{\even}{{\mathrm e}}
\newcommand{\GGE}{{\mathrm{GGE}}}

\def\tit#1{}
\newcommand{\half}{\frac{1}{2}}

\eqnobysec

\begin{document}

\title[Exactly solvable deterministic model of ballistic and diffusive transport]{Exactly solvable deterministic lattice model of crossover between ballistic and diffusive transport}

\author{Katja Klobas, Marko Medenjak, Toma\v z Prosen}
\address{Department of Physics, Faculty of Mathematics and Physics, University of Ljubljana, Ljubljana, Slovenia}

\begin{abstract}
We discuss a simple deterministic lattice gas of locally interacting charged particles, for which we show coexistence of ballistic and diffusive transport. Both, the ballistic and the diffusive transport coefficients,
specifically the Drude weight and the diffusion constant, respectively, are analytically computed for particular set of generalised Gibbs states and may  independently vanish for appropriate values of thermodynamic parameters.
Moreover, our analysis, based on explicit construction of the matrix representation of time-automorphism in a suitable basis of the algebra of local observables, allows for an exact computation of the dynamic structure factor and closed form solution of the inhomogeneous quench problem.
\end{abstract}

\section{Introduction}
One of the main challenges of nonequilibrium statistical mechanics is a rigorous derivation, without any assumptions or approximations, of irreversible macroscopic transport laws, from the microscopic reversible equations of motion.
This goal can be achieved only in certain specific interacting modes, see e.g. a very recent work \cite{bob2}, which can be considered as a companion to the present text.
The aim of this article is to explore dynamical and transport properties of the model of a reversible cellular automaton introduced and preliminary studied in \cite{PhysRevLett.119.110603}, consisting of locally interacting positively
and negatively charged particles and freely propagating vacancies. This work is therefore connected to two active areas of research.
First of all it can be viewed as a discrete-time (paralel-update) and deterministic version of the exclusion
processes~\cite{liggett1975ergodic,Rajewsky1998,VANICAT2018298}, which have been widely
studied~\cite{derrida1993exact,blythe2007nonequilibrium}. In those models
the particles obey exclusion principle, i.e.\ a~particle can move to
a neighboring site only if the site is unoccupied.  Additionally, our model can be
related to the gas of hard rods, which has been extensively studied \cite{spohn2012large,doyon2017dynamics,Spohn1977,lebowitz1968,lebowitz1969,durr1985,VanDenBroeck}, in a specific simple, yet nontrivial regime where the dynamics can be mapped to a spatio-temporal lattice. Due to the model's simplicity many essential questions regarding the dynamics and the transport can be answered explicitly. We hope that these results will illuminate a general understanding of dynamics in a class of similar or related interacting models.

Furthermore, our results can be put into a~more modern perspective of the generalized
hydrodynamics, which was developed
recently~\cite{bertini,PhysRevX.6.041065,doyon2017geometric,ilievski2017microscopic,ilievski2017ballistic,bulchandani2017bethe}
and provides an exact semi-classical description of the long time dynamics of integrable quantum systems.
Semi-classical dynamics is connected to the generalization of the hard
rod gas~\cite{doyon2017soliton}. Taking into account some approximations,
quantum systems can be described in terms of multi-species hard-core colliding
particles~\cite{kormos2017semiclassical,PhysRevB.84.165117,1742-5468-2013-04-P04003,PhysRevE.93.062101,PhysRevLett.119.100603}, exactly the type of model studied here.

In the~article we consider a deterministic dynamics describing
the elastic scattering of charged particles, as well as a Markovian generalization,
where the particles exchange their position with a certain probability upon interaction. We provide analytical expressions
for dynamical (transport) coefficients (specifically, the diffusion constant and the Drude weight) and compare our results to the effective (hydrodynamic)
description, which is typically used without rigorous justification.

In the~first section we introduce the model and the underlying mathematical
structure. Our model supports solitonic excitations. Characterizing the space of  solitonic
observables we are able to construct the set of local conserved
quantities, which describe the stationary states of our model in terms of the
corresponding Generalized Gibbs Ensemble.

The second section deals with the exact calculation of transport coefficients. In
the first part we introduce the linear response definitions of the charge diffusion
constant and the Drude weight. This is followed by the calculation of the lower
bound on Drude weight by employing the Mazur inequality. The central part of
the section comprises the analytical calculation of the linear response diffusion
constant and the Drude weight.

In the third section, a large time solution of the inhomogeneous initial value
problem is obtained, showing ballistic propagation of the step function profile
with diffusive corrections. Depending on the density of vacancies and the
imbalance of particles the system is shown to exhibit ballistic, normal or
isolating behavior. The results following from the inhomogeneous initial state
are compared to the hydrodynamical picture, showing perfect agreement.

In the fourth section we show how the spatio-temporal correlation functions can be
obtained for particular stationary product states. The dynamics corresponds to
the diffusive broadening of the central peak, and the free propagation of
solitary excitations.

\section{The model}
The model is defined on the chain (one-dimensional {\em periodic} lattice) of {\em even} size $n$, where each lattice site can be
occupied by three types of particles: positively charged particles ($+$), negatively
charged particles ($-$) or vacancies ($\emptyset$). The configuration of particles at
time $t$ is denoted by $\underline{s}^{t} $,
with $\underline{s}^{t}=(s^{t}_1,s^{t}_2,...,s^{t}_n)$
and~$s^{t}_x\in\{\emptyset,+,-\}$. The dynamics of particles is described by the~propagation
rule
\begin{eqnarray}
  \eqalign{
    \underline{s}^{t+1}&=\phi( \underline{s}^t), \\
    \phi&=\phi^\odd\circ \phi^\even,\\ 
    \phi^\odd&=\phi_{1,2}\circ \dots \circ \phi_{n-1,n},\\
    \phi^\even&=\phi_{2,3}\circ \dots \circ \phi_{n,1},}
\end{eqnarray}  
comprising two site interactions
\begin{eqnarray}
  \phi_{x,x+1}(\un{s})=(s_1,s_2,\dots,s'_x,s'_{x+1},\dots,s_n),
\end{eqnarray}
that correspond to the following local mapping
\begin{eqnarray}\label{eq:localpropagator}
  \eqalign{
    (s_x,s_{x+1})\leftrightarrow(s_x',s_{x+1}'):\,&(\emptyset,\emptyset) \leftrightarrow  (\emptyset,\emptyset),\\
    &(\emptyset,\alpha)  \leftrightarrow  (\alpha,\emptyset), \\
    &(\alpha,\beta)  \leftrightarrow  (\alpha,\beta).
  }
\end{eqnarray}
\begin{figure}
  \centering
  \definecolor{negative}{RGB}{199,233,180}
\definecolor{negative}{RGB}{179,209,162}
\definecolor{positive}{RGB}{29,145,192}
\definecolor{positive}{RGB}{110,184,214}
\definecolor{empty}{rgb}{1,1,1}
\begin{tikzpicture}[scale=1.5,positiveline/.style={positive,line width=0.75pt},negativeline/.style={negative,line width=0.75pt},stylerec/.style={gray,rounded corners}]
  \node at (0,1) {$t$};
  \node at (0,2) {$t+\frac{1}{2}$};
  \node at (0,3) {$t+1$};

  \node at (1,0.25) {$\emptyset$};
  \node at (2,0.25) {$+$};
  \node at (3,0.25) {$+$};
  \node at (4,0.25) {$-$};
  \node at (5,0.25) {$-$};
  \node at (6,0.25) {$\emptyset$};
  \node at (7,0.25) {$-$};
  \node at (8,0.25) {$-$};

  \node at (1,3.75) {$+$};
  \node at (2,3.75) {$+$};
  \node at (3,3.75) {$\emptyset$};
  \node at (4,3.75) {$\emptyset$};
  \node at (5,3.75) {$-$};
  \node at (6,3.75) {$-$};
  \node at (7,3.75) {$-$};
  \node at (8,3.75) {$\emptyset$};

  \draw[positiveline] (3,1) -- (3.5,1.5) -- (3,2) -- (2,3);
  \draw[positiveline] (2,1) -- (1,2) -- (0.6,2.4);
  \draw[positiveline] (0.6,2.6) -- (1,3);
  \draw[negativeline] (4,1) -- (3.5,1.5) -- (4,2) -- (5,3);
  \draw[negativeline] (5,1) -- (6,2) -- (6.5,2.5) -- (6,3);
  \draw[negativeline] (7,1) -- (7.5,1.5) -- (7,2) -- (6.5,2.5) -- (7,3);
  \draw[negativeline] (8,1) -- (7.5,1.5) -- (8,2) -- (8.4,2.4);

  \draw[black,fill=empty] (1,1) circle(3pt);
  \draw[black,fill=positive] (2,1) circle(3pt);
  \draw[black,fill=positive] (3,1) circle(3pt);
  \draw[black,fill=negative] (4,1) circle(3pt);
  \draw[black,fill=negative] (5,1) circle(3pt);
  \draw[black,fill=empty] (6,1) circle(3pt);
  \draw[black,fill=negative] (7,1) circle(3pt);
  \draw[black,fill=negative] (8,1) circle(3pt);
  \draw[black,fill=positive] (1,2) circle(3pt);
  \draw[black,fill=empty] (2,2) circle(3pt);
  \draw[black,fill=positive] (3,2) circle(3pt);
  \draw[black,fill=negative] (4,2) circle(3pt);
  \draw[black,fill=empty] (5,2) circle(3pt);
  \draw[black,fill=negative] (6,2) circle(3pt);
  \draw[black,fill=negative] (7,2) circle(3pt);
  \draw[black,fill=negative] (8,2) circle(3pt);
  \draw[black,fill=positive] (1,3) circle(3pt);
  \draw[black,fill=positive] (2,3) circle(3pt);
  \draw[black,fill=empty] (3,3) circle(3pt);
  \draw[black,fill=empty] (4,3) circle(3pt);
  \draw[black,fill=negative] (5,3) circle(3pt);
  \draw[black,fill=negative] (6,3) circle(3pt);
  \draw[black,fill=negative] (7,3) circle(3pt);
  \draw[black,fill=empty] (8,3) circle(3pt);

  \foreach \i in {1,...,4}{
    \draw [stylerec] ($(2*\i-1,1)+(-0.3,0.3)$) rectangle ($(2*\i,1)+(0.3,-0.3)$);
    \draw [stylerec] ($(2*\i-1,3)+(-0.3,0.3)$) rectangle ($(2*\i,3)+(0.3,-0.3)$);
  };
  \foreach \i in {1,...,3}{
    \draw [stylerec] ($(2*\i,2)+(-0.3,0.3)$) rectangle ($(2*\i+1,2)+(0.3,-0.3)$);
  };
  \draw [stylerec] ($(1,2)+(-0.3,0.3)$) -- ($(1,2)+(0.3,0.3)$) --
  ($(1,2)+(0.3,-0.3)$) -- ($(1,2)+(-0.3,-0.3)$);
  \draw [stylerec] ($(8,2)+(0.3,0.3)$) -- ($(8,2)+(-0.3,0.3)$) --
  ($(8,2)+(-0.3,-0.3)$) -- ($(8,2)+(0.3,-0.3)$);
\end{tikzpicture}
  \caption{\label{model}
    Schematic representation of the dynamics. Upon meeting, two charged particles scatter elastically and move freely otherwise. Alternatively, the dynamics can be understood as the ballistic propagation of vacancies in the background of charged particles.
  Odd pairs of sites are updated between the~time-slices $ t$
and $ t+\half $, while the~even pairs are propagated in the second half-time step.}
\end{figure}
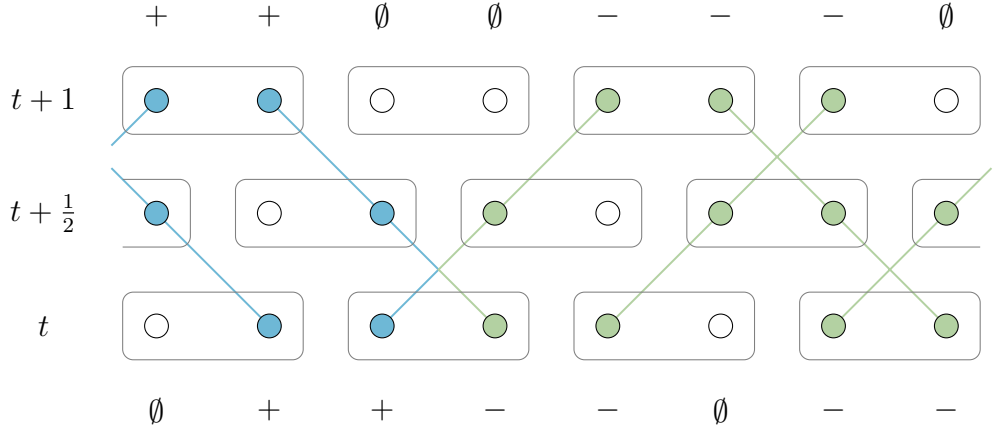%
These rules describe elastic scattering of particles, with scattering in the
first half-time step occurring between the particles on odd-even sites and in
the second half-time step between particles on even-odd sites. The schematic
representation of the dynamics can be seen in Figure~\ref{model}.

The
density of charge~$q^t_x$ is defined as a~sum of net charge on two
neighboring sites,
\begin{eqnarray}\label{gostota}
  q^t_x=s_x^t+s_{x+1}^t,
\end{eqnarray}
where $ s^t_x=1$, if the~site $ (x,t) $ is occupied by a~positive charge, $ -1 $ if
a~negative charge occupies the~site and $ 0 $ in the case of an~empty site. One should
note that the~total charge,
\begin{eqnarray}
  Q=\sum_x q_x,
\end{eqnarray}
is a constant of motion.
To study the dynamics of charges $ q_x $, we introduce the~corresponding current~$j_x$
that satisfies the~continuity equation,
\begin{eqnarray}
  \label{entok}
  j_{x}^{t+1/2}=2 \Big(s^{t+1/2}_{x}-s^{t+1/2}_{x+1}\Big)\Big(s^{t+1/2}_x+s^{t+1/2}_{x+1}\Big)^2.
\end{eqnarray}

\subsection{Algebra of observables}
To describe the statistical properties of the model we
introduce a multiplicative commutative algebra of
observables~$\tilde{{\cal A}} \simeq \mathbb{R}^3 $, i.e. functions over $\mathbb{Z}_3= \{\emptyset,+,-\}$, with the basis $[\alpha]$, 
\begin{eqnarray}
  \eqalign{
  [\alpha](s)=\delta_{\alpha,s}, \qquad \alpha \in  \{\emptyset,+,-\},\\
([\alpha][\beta])(s)= [\alpha](s)\,[\beta](s).}
\end{eqnarray}
The algebra can be extended to a local algebra of functions over the lattice configuration
space~$\left(\mathbb{Z}_{3}\right)^{\otimes n}$ by
defining the~local basis
\begin{eqnarray}
  [\alpha]_x(\ul{s})=\delta_{\alpha_x,s_x}, \qquad \alpha \in  \{\emptyset,+,-\}.
\end{eqnarray}
We introduce the~compact notation for the~local basis elements of
the~tensor product algebra~$\mathcal{A}=\tilde{\mathcal{A}}^{\otimes n}$,
\begin{eqnarray}
  [\alpha_1 \alpha_2 \ldots \alpha_r]_x = [\alpha_1]_{x} [\alpha_2]_{x+1} \cdots [\alpha_r]_{x+r-1}.
\end{eqnarray}
The time propagation of observables can be represented in terms of a linear map $ U\in \text{End}({\cal A})$,
\begin{eqnarray}
  a^t(\underline{s})=a(\underline{s}^t)\equiv U^t a(\underline{s}),
\end{eqnarray}
which is again composed of the~local two site propagators,
\begin{eqnarray}
  \eqalign{
  U = U^\odd U^\even,\\
  U^\odd = \prod_{x=1}^{n/2} U_{2x-1,2x},\\
  U^\even = \prod_{x=1}^{n/2} U_{2x,2x+1}.
}
\end{eqnarray}
Note that lattice sites $1$ and $n+1$ are identified due to periodic boundary conditions.
The two half-time propagators are mutually conjugate, $U^\odd=\eta U^\even \eta^{-1}$, where~$\eta$ is
a lattice shift automorphism defined by $ \eta [\alpha]_x=[\alpha]_{x+1} $, implying
that~$U= \eta U^\even \eta^{-1} U^\even $.
The~expectation value of an~observable~$a$ w.r.t.~the~probability distribution~$p$ over the set of configurations is
\begin{eqnarray}
  \label{probp}
  \langle a\rangle_p=\sum_{\underline{s}} a(\underline{s}) p(\underline{s}).
\end{eqnarray}
Specifically, the~expectation value w.r.t.~the~(non-normalized) maximum entropy state is
\begin{eqnarray}
  \langle a\rangle=\sum_{\underline{s}} a(\underline{s}).
\end{eqnarray}
The expectation value~\eqref{probp} can be represented as $ \langle a\,p \rangle $. For later
convenience we introduce the following two bases for~$\tilde{{\cal A}}$,
depending on the density parameter~$\rho$,
\begin{eqnarray}
  \label{ort_b}
  &[0] =  [\emptyset]+[+]+[-], \qquad &[0]' = (1-\rho) [\emptyset]+\frac{\rho}{2} \left([+]+[-]\right),\nonumber\\
  &[1] =  [+]-[-], &[1]'  =  \frac{1}{2}\left([+]-[-]\right),  \label{basis} \\
  &[2] = \frac{1}{1-\rho}[\emptyset]-[0], &[2]' = \frac{1-\rho}{2}\left(2[\emptyset]-[+]-[-] \right), \nonumber
\end{eqnarray}
which are dual w.r.t.~the maximum entropy state, 
\begin{equation}
\ave{[\alpha][\beta]'} =
\delta_{\alpha,\beta}.
\end{equation} 
The observable~$[0]$ corresponds to the
identity in $\tilde{{\cal A}}$. Let~$\rho=(\rho_1+\rho_2)/2$ be an average
value of the~density parameter on sites~$1$ and~$2$ of a unit cell, comprising two consecutive sites,
while~$\Delta=(\rho_1-\rho_2)/2$ is one half of their~difference.
With respect to this parametrization, the~local propagator takes the following form
\begin{eqnarray}\label{eq:matrixU}
  U_{1,2}=
  \begin{bmatrix}
    1&&&&&&&&\\
    &\rho-\Delta&&\bar{\rho}+\Delta&&\rho-\Delta&&-\rho+\Delta&\\
    &&&&&&1&&\\
    &\bar{\rho}-\Delta&&\rho+\Delta&&-\rho-\Delta&&\rho+\Delta&\\
    &&&&1&&&&\\
    &\bar{\rho}-\Delta&&-\bar{\rho}+\Delta&&\bar{\rho}-\Delta &&\rho+\Delta&\\
    &&1&&&&&&\\
    &-\bar{\rho}-\Delta&&\bar{\rho}+\Delta&&{\rho}-\Delta &&\bar{\rho}+\Delta&\\
    &&&&&&&&1\\
  \end{bmatrix},
\end{eqnarray}
where~$\bar{\rho}=1-\rho$ is a density of vacancies. Note that the propagator is expressed w.r.t.~the basis and the dual basis that have
different values of parameter~$\Delta$, corresponding to the exchange of~$\rho_1$ and~$\rho_2$.
To be more precise, the~elements of the~propagator are defined by
\begin{eqnarray}
  U_{1,2}^{(k,k^{\prime}),(l,l^{\prime})}=\ave{[k,k^{\prime}]^{\prime} U_{1,2}[l,l^{\prime}]},
  \label{U12basis}
\end{eqnarray}
with~$l$ and~$k^{\prime}$ expressed in bases (\ref{basis}) with density~$\rho_1$, and~$l^{\prime}$, $k$ in bases with
density~$\rho_2$.
In these bases the~total charge ($Q$) and the total~current ($J$) observables are expressed as
\begin{eqnarray}
  \fl \label{eq:wholecharge}
  Q =\sum_{x}\left([10]_{2x}\!+\![01]_{2x}\right),\\
  \fl \label{eq:wholecurrent}
    J=\sum_x
    \left(
    (1-\rho_2)\left(
    2\oo{10}_{2x}\!
    \!+\oo{12}_{2x}\!
    \!+\oo{021}_{2x}\right)
    -(1-\rho_1)\left(
    2\oo{01}_{2x}\!
    \!+\oo{21}_{2x}\!
    \!+\oo{012}_{2x}\right)
  \right).
\end{eqnarray}

\subsection{Solitons and local conservation laws}
Local and pseudolocal conservation laws play a key role in the physics local observables \cite{IMPZ,IMP15,Mazur69,Zotos97,PhysRevLett.119.110603}.
Here we show how the~ballistic propagation of vacancies gives rise to {\em exponentially} many local conservation laws, which should be contrasted to expected behaviour in generic integrable systems, where the number of conservation laws scales {\em linearly} with the system size.
First, notice that the~local propagation of~$\oo{20}$ and~$\oo{02}$ is equivalent to the~lattice shift of
observables,~$U_{1,2} \oo{02}=\oo{20}$ and~$U_{1,2}\oo{20}=\oo{02}$.
This is a~consequence of the free propagation of vacancies, and the preservation of the identity observable,
i.e.\ $U_{1,2}\oo{00}=\oo{00}$.

The solitons ensure the~existence of exponentially many conserved quantities, which we now construct.
Let us define the~even solitonic subspace
as
\begin{equation}
\mathcal{S}^{\text{e}}=\lsp\{\oo{2}_{k_1}\!\!\!\cdot \oo{2}_{k_2}\!\cdots \oo{2}_{k_l};\, l\in\{1,2\ldots,\frac{n}{2}\}, \,k_1,
k_2,\ldots,k_l \in 2\ZZ \}
\end{equation} 
and the~odd solitonic subspace 
as
\begin{equation}
\mathcal{S}^{\text{o}}=\lsp\{\oo{2}_{k_1}\!\!\!\cdot \oo{2}_{k_2}\!\cdots \oo{2}_{k_l};\, l\in\{1,2\ldots,\frac{n}{2}\}, \,k_1,
k_2,\ldots,k_l\in 2\ZZ+1 \}.
\end{equation}  
For the elements of these subspaces, the following holds,
\begin{eqnarray}
  U s_{\underline{k}}=\begin{cases}\eta^{2}(s_{\underline{k}}); &s_{\underline{k}}\in\mathcal{S}^\text{e}\\
    \eta^{-2}(s_{\underline{k}});&s_{\underline{k}}\in\mathcal{S}^\text{o}
  \end{cases},
\end{eqnarray}
where~$\underline{k}=(k_1,k_2,\ldots,k_l)$ denotes the places where the~observables~$[2]$ appear.
Therefore any solitonic observable~$s_{\underline{k}}\in \mathcal{S}^{\text{e/o}}$, is a~local density of
a~conserved quantity~$S_{\underline{k}}=\sum_{x}\eta^{2x}(s_{\underline{k}})$,
\begin{eqnarray}\label{eq:solitoniccharges}
  U\,S_{\underline{k}}=U\left(\sum_{x=0}^{n/2-1}
  \eta^{2x}(s_{\underline{k}})\right)=\sum_{x=0}^{n/2-1} \eta^{2x\pm2}
  (s_{\underline{k}})=S_{\underline{k}};\qquad s\in \mathcal{S}^\text{e/o}.
\end{eqnarray}
If the length of the chain is~$n$, the number of distinct solitonic conserved
quantities~$S_{\underline{k}}$ is $2^{n/2}$, i.e.\ square-root of the number of all linearly independent observables.  
As already mentioned, the total
charge~$Q$ is conserved as well. Therefore the~number of all local integrals of
motion is~$2^{n/2}+1$.

\subsection{Generalized Gibbs Ensemble}
A~generic integrable system equilibrates to the non-thermal equilibrium state, which
is expected to be described by the Generalized Gibbs ensemble~\cite{Rigol_PRL07,Ilievski_GGE,ilievski2017interacting}.
Let us consider a~homogeneous initial value problem with a translationally invariant initial state~$\rho_i$, $\eta^2 \rho_i = \rho_i$.
Then for any local observable~$a$ the~infinite time average corresponds to the~ensemble
average,
\begin{eqnarray}
  \label{GGE}
  \lim_{T\to\infty}\frac{1}{T}\sum_{t=0}^T\langle a^t
  \rangle_{\rho_i}=\langle a \rangle_{\rho_{\GGE}}.
\end{eqnarray}
Provided that the complete set of local charges
is~$\{S^{\even}_{\underline{k}},\ S^{\odd}_{\underline{k}},\ Q\} $, the hypothetic equilibrium ensembles can be described by 
\begin{eqnarray}
  \rho_{\GGE}=\frac{1}{Z}\exp\left(-\sum_{\underline{k}}  \beta^\even_{\underline{k}}
  S^\even_{\underline{k}}-\sum_{\underline{k}} \beta^\odd_{\underline{k}}
  S^\odd_{\underline{k}}-\beta Q\right).
\end{eqnarray} 
Here the chemical potentials~$\beta,\beta^{\even/\odd}_{\underline{k}}$ are obtained from the condition that the
initial state expectation values of conserved charges matches the ones in the
equilibrium state~$\rho_{\GGE}$. The expression for~$\rho_{\GGE}$ can be
further simplified,
\begin{eqnarray}
  \label{GGE2}
  \rho_{\GGE}=\Big(1-\sum_{\underline{k}}
  c^\even_{\underline{k}}S^\even_{\underline{k}}\Big)\Big(1-\sum_{\underline{k}}
  c^\odd_{\underline{k}}S^\odd_{\underline{k}}\Big) p,
\end{eqnarray}
where we introduced the~product equilibrium distribution~$p$,
\begin{eqnarray}\label{GGEp}
  p=\prod_x\Big(([0]_{2 x}'+\mu_1 [1]_{2 x}')([0]_{2 x+1}'+\mu_2 [1]_{2 x+1}')\Big).
\end{eqnarray}
Note that here the dual basis vectors pertaining to even and odd sites correspond
to~$ \rho_1 $ for even and to~$ \rho_2 $ for odd sites,
however, we suppressed the explicit dependence to preserve the compactness of notation.
The first two terms in equation~\eqref{GGE2} are obtained by expanding
the~exponents of even and odd charges,
and noting that the~exponents of even/odd solitonic observables are also
even/odd solitonic observables.
The expression~\eqref{GGEp} is obtained by expanding~$ \exp(-\alpha^\even S_1^\even-
\alpha^\odd S_1^\odd-\beta Q) $, and choosing~$ \alpha^\even $ and~$ \alpha^\odd $ such that
the expansion of the exponent does not contain basis vectors~$ [2]' $.  This
imposes the following restriction,
\begin{eqnarray}
  \frac{\mu_1}{\mu_2}=\frac{\rho_1}{\rho_2}.
\end{eqnarray}
For the purpose of
hydrodynamic applications we are interested only in expectation values of
the observables from the
set $\mathcal{J}=\{s_{\underline{k}}^{\text{e/o}},q;j_{\underline{k}}^{\text{e/o}},j\}$,
which includes densities of conserved quantities and their currents.
The only current that is not a linear combination of charges is the charge
current $ j $. On the subspace of charges and
currents, the state \eqref{GGE2} can be equivalently represented as
\begin{eqnarray} \label{GGEt}
  \tilde{\rho}_{\GGE}=\bigg(1-\sum_{\underline{k}}\left(c^+_{\underline{k}}
  S^+_{\underline{k}}+c^-_{\underline{k}} S^-_{\underline{k}}\right) \bigg) p,
\end{eqnarray}
where we introduced symmetrized and anti-symmetrized
charges~$ S_{\underline{k}}^\pm=S_{\underline{k}}\pm \eta_1(S_{\underline{k}})$.
Here, we take into account that, for
every~$a\in\mathcal{J}$, the expectation value~$\ave{a\,S_k^{\text{e}}S_k^{\text{o}}}_p$
vanishes, implying~$\ave{a}_{\rho_{\GGE}}= \ave{a}_{\tilde{\rho}_{\GGE}}$.

In the rest of the article, the~results are obtained for the~factorizable ({\em separable}) steady states~$\rho_{\GGE}=p$~\eqref{GGEp},
i.e. considering $c^{\pm}_{\underline{k}}=0$. The~overlaps of observables w.r.t.~states~$p$ are easy to
express due to the~following relations,
\begin{eqnarray}\label{eq:orthogonal}
  \eqalign{
    \ave{\oo{0}_{2x}\oo{0}_{2x}}_p&=1,\\
    \ave{\oo{0}_{2x}\oo{1}_{2x}}_p&=\mu_1,\\
    \ave{\oo{0}_{2x}\oo{2}_{2x}}_p&=0,\\
    \ave{\oo{1}_{2x}\oo{1}_{2x}}_p&=\rho_1,\\
    \ave{\oo{1}_{2x}\oo{2}_{2x}}_p&=-\mu_1,\\
    \ave{\oo{2}_{2x}\oo{2}_{2x}}_p&=\frac{\rho_1}{1-\rho_1},\\
  }\qquad
  \eqalign{
    \ave{\oo{0}_{2x+1}\oo{0}_{2x+1}}_p&=1,\\
    \ave{\oo{0}_{2x+1}\oo{1}_{2x+1}}_p&=\mu_2,\\
    \ave{\oo{0}_{2x+1}\oo{2}_{2x+1}}_p&=0,\\
    \ave{\oo{1}_{2x+1}\oo{1}_{2x+1}}_p&=\rho_2,\\
    \ave{\oo{1}_{2x+1}\oo{2}_{2x+1}}_p&=-\mu_2,\\
    \ave{\oo{2}_{2x+1}\oo{2}_{2x+1}}_p&=\frac{\rho_2}{1-\rho_2}.
  }
\end{eqnarray}

\section{Linear response}
One of the outstanding questions when regarding the transport phenomena is how
to derive the phenomenological transport laws, e.g. Fick's law, stating that the current
is proportional to the gradient of the external field $\nabla h$,
\begin{eqnarray}
  j=\sigma \nabla h,
\end{eqnarray}
where the proportionality constant~$\sigma$ is the conductivity.
The transport coefficients in the linear response regime can be calculated in
terms of the stationary state time-correlation functions.
The~\emph{diffusion constant}~$\mathcal{D}$ is related to the conductivity through the Einstein's relation,
\begin{eqnarray}
  \sigma=\chi \mathcal{D},
\end{eqnarray}
where~$\chi$ is the static susceptibility. The conductivity
can be expressed as the time integral of the current autocorrelation function,
\begin{eqnarray}
  \sigma=\frac{1}{2}\sum_{t=-\infty}^\infty C_J(t),\qquad
  C_J(t)=\lim_{n\to\infty}\frac{1}{n}\ave{J(t)J}_{\rho_{\GGE}}.
\end{eqnarray}
The \emph{Drude weight}, defined as
\begin{eqnarray}
  D=\lim_{t\to\infty }C_J(t),
\end{eqnarray}
is the rate at which the current in the system increases
when the system is exposed to the constant gradient external field. For precise definitions and derivation
see~\ref{app:linearResponse}.
If the Drude weight is non-zero, the diffusion constant diverges. In this case
it is convenient to regularize it by subtracting the divergent part,
\begin{eqnarray}
  \mathcal{D}=\frac{1}{2}\sum_{t=-\infty}^\infty (C_J(t)-D).
\end{eqnarray}

\subsection{Mazur bound on Drude weight}
In this subsection, we bound the Drude weight by local conservation laws,
generalizing Mazur's argument~\cite{Mazur69} to the discrete time case.
Let us consider a~time-averaged observable $\bar{a}= \frac{1}{T}\sum_{t=0}^T a^t $.
The~following inequality holds for
any stationary probability distribution~$p$,
\begin{eqnarray}
  \label{ineq}
  \langle \bar{a}^2 \rangle_p\geq 0.
\end{eqnarray}
Mazur's inequality then follows directly 
by setting~$a=\frac{1}{n}(J-\sum_k \alpha_k O_k) $, where~$\{O_k\}$ is a set
of conserved charges. Inserting~$\bar{a} $ into the~expression~\eqref{ineq},
and maximizing the~expression w.r.t.~the~set~$\{\alpha_k\}$ one obtains the
following lower bound,
\begin{eqnarray}\label{eq:mazurinequality}
  \lim_{T\to\infty}\frac{1}{2 T}\sum_{t=-T}^T \frac{1}{2\,n}\langle J(t)J\rangle_p\geq
  \frac{1}{2\, n}\sum_{k} \frac{\langle J O_k\rangle_p^2}{\langle O_k^2\rangle_p}.
\end{eqnarray}
Here we assumed that the~set of the~charges is orthogonal, i.e.\ $\langle O_k\, O_l  \rangle_p=0 $.

Let us now proceed to the calculation of the Mazur bound.
The only two solitonic charges that have nonzero overlap with
the~current are
\begin{eqnarray}
  S_{(1)}^{\text{e}}=\sum_x \eta_{2x}(\oo{02}),\qquad
  S_{(1)}^{\text{o}}=\sum_x \eta_{2x}(\oo{20}).
\end{eqnarray}
Additionally, the following linear combination of~$Q$ and~$S_{(1)}^{\text{e/o}}$ also contributes
to the~Mazur bound,
\begin{eqnarray}\label{conservedquantityo0}\fl
  \tilde{S}=\sum_x \eta_{2x}\bigg(\oo{01}+\oo{10}+\frac{\mu}{\rho}(1-\rho)\left(\oo{20}+\oo{02}\right)+
  \frac{\mu}{\rho}\Delta \left(\oo{02}-\oo{20}\right)\bigg).
\end{eqnarray}
Therefore we are able to utilize Mazur's inequality~\eqref{eq:mazurinequality}, and bound
the~current autocorrelation function,
\begin{eqnarray}
\lim_{T\to\infty}\frac{1}{2 T}\sum_{t=-T}^T \frac{1}{2\,n}\langle J(t)J\rangle_p\ge
  \frac{1}{2\, n}\left(
    \frac{\langle J \tilde{S}\rangle_p^2}{\langle \tilde{S}^2\rangle_p}+
    \frac{\langle J S_{(1)}^{\text{e}}\rangle_p^2}{\langle S_{(1)}^{\text{e}\,2}\rangle_p}+
  \frac{\langle J S_{(1)}^{\text{o}}\rangle_p^2}{\langle S_{(1)}^{\text{o}\,2}\rangle_p}\right),
\end{eqnarray}
yielding the following lower bound on the Drude weight,
\begin{eqnarray}\label{eq:lowerbound}
  \eqalign{
    D &= \lim_{t\to\infty}\lim_{n\to\infty}
    \frac{1}{t}
    \left(\frac{1}{2n}\ave{J^2}_p+
    \sum_{t^{\prime}=1}^t\frac{1}{n}\ave{J(t^{\prime}) J}_p\right)\ge\\
    &\ge
    16\frac{1-\rho}{\rho}\mu^2+16\frac{\Delta^2}{\rho}\left(1-\frac{\mu^2}{\rho^2}(1+\rho)\right).
  }
\end{eqnarray}
In the following subsection we show that the lower bound saturates the exact result.

\subsection{Exact results on time decay of current-current autocorrelation}
The main property of the time-dependence of the autocorrelation
functions is that time propagation can be restricted to a particular subspace,
where the propagator has at most five-diagonal form. This allows for an
explicit calculation of~$C_J(t)$. 

In order to calculate the~autocorrelation we first observe
that the number of local basis elements~$\oo{1}$, occurring in the time
propagated observables, is conserved. Furthermore, the basis elements~$\oo{2}$
propagate ballistically, and due to the~orthogonality~\eqref{eq:orthogonal}
we can restrict the computation of current-current autocorrelation function, $C_J(t)$,
to the
subspace~$\mathcal{A}_J=\text{lsp}\{y^+_0,y^-_0,y^+_1,y^-_1,z^+_{2d,0},z^-_{2d,0},z^+_{2d+1,1},z^-_{2d+1,1};\,d\ge 0\}$,
with the basis elements~$y_0^{\pm}$, $y_1^{\pm}$, $z_{d,k}^{\pm}$ defined as
\begin{eqnarray}
  \eqalign{
    y^{\pm}_0 &= \sum_x \oo{10}_{2x}\pm \oo{01}_{2x}, \\ 
    y^{\pm}_1 &= \sum_x \oo{12}_{2x}\pm \oo{21}_{2x}, \\
    z_{d,k}^{\pm} &=\sum_x\oo{0 \underbrace{0 \ldots 0}_{k} 1 \underbrace{0\ldots 0}_d 2}\vphantom{\oo{21}}\emptysub{2x}
    \pm\oo{0 2\underbrace{0\ldots 0}_d 1 \underbrace{0\ldots 0}_k}\emptysub{2x}.
  }
\end{eqnarray}
Formally, the argument can be recast as follows. We can consider only observables with a~single~$\oo{1}$, i.e.\ those spanned by any local basis vectors $[\alpha_1 \alpha_2 \cdots \alpha_r]_x$ with a single $\alpha_i = 1$,
forming an invariant subspace~$\mathcal{A}^{(1)}$, $U \mathcal{A}^{(1)} \subseteq \mathcal{A}^{(1)}$.
Let~$P$ be a~linear projector~$P:\,\mathcal{A}^{(1)}\to \mathcal{A}_J$,
$P^2=P$. Then for every $a\in\mathcal{A}^{(1)}$, $\ave{J (1-P) a}_p=0$,
and $\mathcal{A}^{(1)}/\mathcal{A}_J$ is invariant under $\eta U^{\text{e}}$ and $\eta^{-1} U^{\text{e}}$
This implies that in order to compute the current autocorrelation function the dynamics can be restricted to~$\mathcal{A}_J$.
Specifically, by defining the reduced half-time step propagator~$\mathcal{U}$,
\begin{eqnarray}
  \mathcal{U}=P\,\eta U^{\text{e}}P = P\,\eta^{-1} U^{\text{e}}P,
\end{eqnarray}
it is possible to express the current-current autocorrelation function exactly as
\begin{eqnarray}\label{eq:reducedautocorrelation}
  C_J(t)=\ave{J\,\mathcal{U}^{2t} J}_p.
\end{eqnarray}
Note that if we consider an initial state with $ \rho_1\neq \rho_2 $, the basis
vectors are position dependent, according to~\eqref{ort_b}. Initially the basis on even sites corresponds to the density of particles $ \rho_1 $ and on odd sites to the density $ \rho_2 $, however after every half-time step the two bases get exchanged.

The reduced propagator reads
\begin{eqnarray}
  \setcounter{MaxMatrixCols}{20}
  \mathcal{U}=
  \begin{tikzpicture}[baseline=(current bounding box.center),scale=1.1, every node/.style={scale=0.9,transform shape},
      every left delimiter/.style={xshift=2mm},
      every right delimiter/.style={xshift=-2mm},
    ]
    \matrix (M)[matrix of math nodes, nodes in empty cells,
      left delimiter={[}, right delimiter={]},
    ]{
      1 & 2\,\Delta & 0 & -2\Delta & 0&0&\vphantom{0}\hphantom{1-\Delta}& & &&\\
      0& 1-2\rho & 0 & 2\rho & 0&0&\vphantom{\ddots}& & &&\\
      0& 0 & 0 & 0 & 0&0&\vphantom{\ddots}& & &&\\
      0& 0 & 0 & 0 & 0&0&\vphantom{\ddots}& & &&\\
      0& 2\Delta & 1 & -2\Delta & 0 & 0&\vphantom{\ddots}& & &&\\
      0& -2(1-\rho) & 0 & 1-2\rho & 0 & 0&\vphantom{\ddots}& & &\phantom{0}&\phantom{\ddots}\\
      & & & &\rho&-{\Delta}&1-\rho&{\Delta}&0&0&\phantom{\ddots}\\
      & & & &-{\Delta}&\rho &{\Delta}&1-\rho&0&0&\phantom{\ddots}\\
      & & & &1-\rho&-{\Delta}&\rho&{\Delta}&0&0&\phantom{\ddots}\\
      & & & &-{\Delta}&1-\rho&{\Delta}&\rho&0&0&\phantom{\ddots}\\
      & & & & & &&&&&\phantom{\ddots}\\
    };
    \draw[densely dashed,thick]([yshift=-0.7mm]M-6-1.south west)--([yshift=-0.7mm]M-6-11.south west);
    \draw[densely dashed,thick]([yshift=0.65mm]M-10-6.south east)--(M-1-7.north west);
    \draw[densely dashed,thick]([yshift=0.65mm]M-10-6.south east)--([yshift=0.65mm]M-11-11.north east);
    \draw[densely dashed,thick](M-7-11.north west)--(M-11-11.south west);
    \begin{scope}[on background layer]
      \draw[draw opacity=0,fill=red, fill opacity=0.2] (M-1-1.north west) rectangle (M-6-4.south east);
      \draw[draw opacity=0,fill=green, fill opacity=0.2] (M-6-4.south east) rectangle (M-10-9.south west);
      \draw[draw opacity=0,fill=green, fill opacity=0.2] (M-10-9.south west) rectangle (M-11-11.south east);
    \end{scope}
  \end{tikzpicture}.
\end{eqnarray}
The~green $ 4\times 4 $ blocks are repeating, and are shifted by 2 columns to the left of the diagonal.
Additionally, we introduce the~vector of overlaps~$\underline{o}$ and
the~appropriately rescaled current vector~$\underline{J}$,
\begin{eqnarray}
\underline{J}=\begin{bmatrix}
\frac{2\Delta}{1-\rho}\\
2\\
\frac{\Delta}{1-\rho}\\
1\\
\frac{\Delta}{1-\rho}\\
\tikzmark{firstline}{-1}\\
0\\
0\\
0\\
\tikzmark{secondline}{0}\\
0\\
0\\
0\\
\tikzmark{thirdline}{0}\\
\vdots
\end{bmatrix},\qquad
\underline{o}=\begin{bmatrix}
2\Delta\\
2\Delta^2+2\rho(1-\rho)\\
0\\
{\rho^2}-{\Delta^2}\\
0\\
\tikzmark{fourthline}{-\left(\rho^2-\Delta^2\right)}\\
0\\
0\\
0\\
\tikzmark{fifthline}{0}\\
0\\
0\\
0\\
\tikzmark{sixthline}{0}\\
\vdots
\end{bmatrix}
+\mu^2
\begin{bmatrix}
-4\frac{\Delta}{\rho}\\
-4\frac{\Delta^2}{\rho^2}\\
0\\
1-\frac{\Delta^2}{\rho^2}\\
0\\
\tikzmark{seventhline}{-\left(1-\frac{\Delta^2}{\rho^2}\right)}\\[7pt]
-4\frac{\Delta}{\rho}\\
-2\left(1+\frac{\Delta^2}{\rho^2}\right)\\
0\\
\tikzmark{eithline}{-2\left(1-\frac{\Delta^2}{\rho^2}\right)}\\[7pt]
-4\frac{\Delta}{\rho}\\
-2\left(1+\frac{\Delta^2}{\rho^2}\right)\\
0\\
\tikzmark{ninthline}{-2\left(1-\frac{\Delta^2}{\rho^2}\right)}\\[7pt]
\vdots
\end{bmatrix}.%
\tikz[overlay,remember picture]{
	\draw[thick,densely dashed] ([xshift=-0.4mm]firstline.south west) -- ([xshift=0.4mm]firstline.south east);
	\draw[thick,densely dashed] ([xshift=-2.05mm]secondline.south west) -- ([xshift=2.05mm]secondline.south east);
	\draw[thick,densely dashed] ([xshift=-2.05mm]thirdline.south west) -- ([xshift=2.05mm]thirdline.south east);
	
	\draw[thick,densely dashed] ([xshift=-3.9mm,yshift=0.75mm]fourthline.south west)
	-- ([xshift=3.9mm,yshift=0.75mm]fourthline.south east);
	\draw[thick,densely dashed] ([xshift=-13.3mm]fifthline.south west) -- ([xshift=13.3mm]fifthline.south east);
	\draw[thick,densely dashed] ([xshift=-13.3mm]sixthline.south west) -- ([xshift=13.3mm]sixthline.south east);
	
	\draw[thick,densely dashed] ([xshift=-1mm,yshift=0.65mm]seventhline.south west) -- ([xshift=1mm,yshift=0.65mm]seventhline.south east);
	\draw[thick,densely dashed] ([yshift=0.65mm]eithline.south west) -- ([yshift=0.65mm]eithline.south east);
	\draw[thick,densely dashed] ([yshift=0.65mm]ninthline.south west) -- ([yshift=0.65mm]ninthline.south east);
},
\end{eqnarray}
so that
the autocorrelation function~\eqref{eq:reducedautocorrelation} reduces to
\begin{eqnarray}
  C_J(t)=16(1-\rho) \underline{o}^{T}\mathcal{U}^{2t}\underline{J}.
\end{eqnarray}
Applying the matrix $ \mathcal{U} $ to the left we obtain an explicit expression for the time dependent correlation function
\begin{eqnarray}
  \fl\nonumber
  C_J(0)=8(1-\rho)(\mu^2+\rho(2-\rho))+8\Delta^2\left(3-\rho-\frac{\mu^2}{\rho^2}(5-\mu)\right),\\
  \fl\label{eq:fullautocorrelation}
  C_J(t>0)=
  16\rho\bigg(
    \frac{\mu^2}{\rho^2}\left(1-\rho\right)
    +\frac{\Delta^2}{\rho^2}\left(1-\frac{\mu^2}{\rho^2}(1+\rho)\right)
  \bigg)+
  \\\nonumber
  +16\rho
    (1-\rho)^4
    \left(1-\frac{\mu^2}{\rho^2}\right)
    \left(1-\frac{\Delta^2}{\rho^2}\right)
    (1-2\rho)^{2t-2}.
\end{eqnarray}

If $ \Delta=0 $ and $ \mu=0 $, i.e.\ for a~translationally invariant state
without the~charge imbalance, the Drude weight vanishes and
the transport is diffusive with the~following diffusion constant
and conductivity,
\begin{eqnarray}\label{eq:realDiffusionConstant}
  \mathcal{D}=\rho^{-1}-1,\qquad
  \sigma=4(1-\rho).
\end{eqnarray}
Otherwise the transport is ballistic, and the Drude weight reads
\begin{eqnarray}\label{eq:wholedrudeweight}
  D=16\,\mu^2\left(\rho^{-1}-1\right)
  +16\,\Delta^2 \rho^{-1}\left(1-\frac{\mu^2}{\rho^2}\left(1+\rho\right)\right),
\end{eqnarray}
which is equal to the~Mazur bound~\eqref{eq:lowerbound}.

\subsection{Stochastic generalization}
Similar calculation can be repeated for a~stochastic generalization,
which corresponds to the tunneling of the particles, by allowing two additional processes
\begin{eqnarray}
  (\pm,\mp)\leftrightarrow(\mp,\pm).
\end{eqnarray}
Let~$\Gamma$ and~$\bar{\Gamma}$ denote the tunneling and
scattering probabilities respectively, namely $(\pm,\mp)$ maps to $(\mp,\pm)$ with probabiliy $\Gamma$, and to $(\pm,\mp)$ with probability $\bar{\Gamma}=1-\Gamma$.
In this case the local propagator (\ref{U12basis}) in the basis (\ref{basis}) reads
\begin{eqnarray}\fl
  U_{1,2}=
  \begin{bmatrix}
    1&&&&&&&&\\
    &\rho_2\bar{\Gamma}&&1-\rho_2 \bar{\Gamma}&&\rho_2\bar{\Gamma}&&-\rho_2\bar{\Gamma}&\\
    &&&&&&1&&\\
    &1-\rho_1\bar{\Gamma}&&\rho_1\bar{\Gamma}&&-\rho_1\bar{\Gamma}&&\rho_1\bar{\Gamma}&\\
    &&&&1&&&&\\
    &(1-\rho_1)\bar{\Gamma}&&-(1-\rho_1)\bar{\Gamma}&&(1-\rho_1)\bar{\Gamma}&&1-(1-\rho_1)\bar{\Gamma}&\\
    &&1&&&&&&\\
    &-(1-\rho_2)\bar{\Gamma}&&(1-\rho_2)\bar{\Gamma}&&1-(1-\rho_2)\bar{\Gamma}&&(1-\rho_2)\bar{\Gamma}&\\
    &&&&&&&&1\\
  \end{bmatrix}.
\end{eqnarray}
Due to the change of dynamics which is no longer deterministic, the~current has to be redefined in order for the~continuity equation
to hold (see \ref{app:linearResponse} for the details), 
\begin{eqnarray}\fl
    J=\sum_x\bigg(\nonumber
      \left(1+\frac{\bar{\Gamma}^2\rho_2}{1-2\bar{\Gamma}}\right)\oo{10}_{2x}
      +\frac{\bar{\Gamma}(1-\rho_2)}{2}\left(
        -\frac{1}{1-2\bar{\Gamma}}\oo{12}_{2x}+\oo{021}_{2x}
      \right)-\\
      -\left(1+\frac{\bar{\Gamma}^2\rho_1}{1-2\bar{\Gamma}}\right)\oo{01}_{2x}
      -\frac{\bar{\Gamma}(1-\rho_1)}{2}\left(
      -\frac{1}{1-2\bar{\Gamma}}\oo{21}_{2x}+\oo{012}_{2x}
      \right)
    \bigg).
\end{eqnarray}
Most of the discussion corresponding to the deterministic dynamics still applies and results in an explicit expression for the time-dependent correlation function,
\begin{eqnarray}\fl\label{eq:stochautocorr}
  C_J(t) =\ave{J^{\frac{1}{2}}J^{t+\frac{1}{2}}}
  =\ave{ (U^{\rm o} J) (U^{\rm o} U^t J)} \nonumber \\ \fl
  =
  16\rho\left(
    \frac{\mu^2}{\rho^2}\left(1-\rho\right)
    +\frac{\Delta^2}{\rho^2}\left(1-\frac{\mu^2}{\rho^2}(1+\rho)\right)
  \right)+
  \\\nonumber\fl
  +16\rho
   \left(1-2\bar{\Gamma}\rho\right)^{2t-2}
  \left(1-\bar{\Gamma}\rho\right)^2
    \left(1-\frac{\mu^2}{\rho^2}\right)
    \left(1-\frac{\Delta^2}{\rho^2}\right)
    \left(1-\bar{\Gamma}\left(3-2\bar{\Gamma}\right)\rho\right)+ \\ \nonumber \fl
    + 4 \Gamma \Bar{\Gamma}^3 \rho^3 (1-\rho) \left(1-\frac{\mu^2}{\rho^2}\right)\left(1-\frac{\Delta^2}{\rho^2}\right)\times
    \\\nonumber
    \times\bigg\{
      \left(2-\bar{\Gamma}(\rho-\Delta)\right)^2\left(1-\frac{\Delta}{\rho}\right)\left(1-\frac{\Delta}{1-\rho}\right)
      \left(1-\alpha(\rho-\Delta)\right)^{2t-2}+\\\nonumber
      \quad\!+\left(2-\bar{\Gamma}(\rho+\Delta)\right)^2\left(1+\frac{\Delta}{\rho}\right)\left(1+\frac{\Delta}{1-\rho}\right)
      \left(1-\alpha(\rho+\Delta)\right)^{2t-2}
    \bigg\}\,.
\end{eqnarray}

Similarly as in the~deterministic case, the~transport is ballistic if~$\mu\neq 0$ or~$\Delta\neq 0$,
with the~same Drude weight, while in the regime~$\mu=\Delta=0$,
the transport is diffusive.
Setting~$\mu=\Delta=0$ in \eqref{eq:stochautocorr}
and evaluating the~sum~\eqref{tk2} yields the~following expression for the diffusion constant
\begin{eqnarray}\fl
  \eqalign{
  &\mathcal{D}=\left(\bar{\rho}^{-1}-1\right) -\\
  &\phantom{\mathcal{D}}- 4 \Gamma + 9\Gamma \bar{\rho}+2\Gamma(4-9\Gamma)\bar{\rho}^2
  -2\Gamma(15-8\Gamma) \bar{\rho}^3+4\Gamma(5-\Gamma) \bar{\rho}^4 -4\Gamma \bar{\rho}^5,
}
\end{eqnarray}
where~$\bar{\rho}$ is the~rescaled density,~$\bar{\rho}=\bar{\Gamma}\rho$.
In the deterministic limit~$\Gamma\to0$, the~diffusion constant of the deterministic model is recovered, while for nonzero
values of the~scattering probability,
$\mathcal{D}$ scales as~$\left(\bar{\Gamma}\rho\right)^{-1}-1$ with polynomial corrections.

\section{Inhomogeneous quench}\label{sec:inhomogeneousQuench}
In this section we consider an inhomogeneous quench problem, where the~initial
probability distribution is given by the product state with the~density of particles $\rho_1$
on odd sites, and $ \rho_2 $ on even sites. Additionally we set a fixed
expectation value of the charge $ q $ to $ \mu_{1,2}^{\rm L/R} $,
with the~superscript indices denoting the~left and the right half of the chain, while
the~indices~$1$ and $2$ correspond to odd and even sites respectively. Note that we are considering the cases where the system size is divisible by $ 4 $, and the local propagation is initially applied to the sites $ (1,2),(3,4),... $
The initial state probability distribution then reads
\begin{eqnarray}\fl
  p=\prod_{x=1}^{n/4}(\oo{0 0}^{\prime}_{2x-1}+\mu_1^{\rm L} \oo{10}^{\prime}_{2x-1}+\mu_2^{\rm L} \oo{01}^{\prime}_{2x-1})
  \!\!\!\!\!\prod_{x=n/4+1}^{n/2}\!\!\!\!\!(\oo{0 0}^{\prime}_{2x-1}
  +\mu_1^{\rm R} \oo{10}^{\prime}_{2x-1}+\mu_2^{\rm R} \oo{01}^{\prime}_{2x-1}).
\end{eqnarray}

The objective is to compute the steady state profile of the~charge~$q$,
\begin{eqnarray}\label{eq:chargeProfileDef}
  f(x,t)=\langle q_{2x}\cdot p^t  \rangle.
\end{eqnarray}
Since the number of local states $\oo{1}^{\prime}$ is preserved,
the state $p$ can be linearized,
\begin{eqnarray}
  \tilde{p}=\sum_{x=1}^{n/4}(\mu_1^{\rm L} \oo{10}^{\prime}_{2x-1}+\mu_2^{\rm L} \oo{01}^{\prime}_{2x-1})
  +\!\!\!\!\sum_{x=n/4+1}^{n/2}\!\!\!\!(\mu_1^{\rm R} \oo{10}^{\prime}_{2x-1}+\mu_2^{\rm R} \oo{01}^{\prime}_{2x-1}).
\end{eqnarray}
Similarly as before we can consider the subspace with a~single local
state~$\oo{1}^{\prime}$. The half-step propagation
on this subspace is
\begin{eqnarray}
  \eqalign{
    &[0 1]_{2 x}^{\prime}
  \mapsto \overline{\Gamma} \rho_1 [0 1]_{2 x}^{\prime}+(1-\overline{\Gamma}\rho_1)[1 0]_{2 x}^{\prime},\\
    &[1 0]_{2 x}^{\prime}
\mapsto  \overline{\Gamma} \rho_2 [1 0]_{2 x}^{\prime}+(1-\overline{\Gamma}\rho_2)[0 1]_{2 x}^{\prime},}
\end{eqnarray}
where~$\overline{\Gamma}=1-\Gamma$.
Note that after the half-time step the change of basis occured
(i.e.\ $ \rho_1\leftrightarrow\rho_2 $).
Let us introduce the~basis of the~linear space spanned by the~local charge
densities,
\begin{eqnarray}
  \hat{e}_x=\oo{1}_x'.
\end{eqnarray}
In this basis the~full time step takes the~form 
\begin{eqnarray}
  \mathcal{U}&=\begin{bmatrix}
    &\ddots&\phantom{\ddots}&&\\
    &C&A&B&&\\
    &&C&A&B&\\
    &&&\phantom{\ddots}&\ddots&\\
  \end{bmatrix},
\end{eqnarray}
with the~$2\times 2$ blocks~$A$, $B$, $C$ given by~
\begin{eqnarray}
  \eqalign{
  A&=
  \begin{bmatrix}
    \overline{\Gamma}^2\rho_1 \rho_2  & \left(1-\overline{\Gamma}\rho_1\right) \overline{\Gamma}\rho_1 \\
    \left(1-\overline{\Gamma}\rho_2\right) \overline{\Gamma}\rho_2 &  \overline{\Gamma}^2\rho_1 \rho_2 \\
  \end{bmatrix},\\
  B&=
  \begin{bmatrix}
    0 & 0 \\
    \overline{\Gamma} (1 - \overline{\Gamma} \rho_1) \rho_2 &
    (1 - \overline{\Gamma} \rho_1)^2  \\
  \end{bmatrix},\\
  C&=
  \begin{bmatrix}
    (1 - \overline{\Gamma} \rho_2)^2 & \overline{\Gamma} (1 - \overline{\Gamma} \rho_2) \rho_1 \\
    0 & 0 \\
  \end{bmatrix},}
\end{eqnarray}
while the initial state can be expressed as
\begin{eqnarray}
  \underline{p}=
  \begin{bmatrix}
    \mu_1^{\rm L}&\mu_2^{\rm L}&\ldots&
    \mu_1^{\rm L}&\mu_2^{\rm L}&\mu_1^{\rm R}&\mu_2^{\rm R}&\ldots&\mu_1^{\rm R}&\mu_2^{\rm R}
  \end{bmatrix}.
\end{eqnarray}
In what follows, we restrict the~discussion to the~deterministic case ($\Gamma=0$, $\overline{\Gamma}=1$), since
the~results for the~stochastic generalization can be obtained 
by rescaling~$\rho_{1,2}\to \overline{\Gamma} \rho_{1,2} $.

The matrix~$ \mathcal{U} $ can be diagonalized using the~block Fourier transform, namely writing the eigenvector with eigenvalue $\lambda(k)$ in the form
\begin{eqnarray}
\bigoplus_r \e^{\ii k r}  \underline{v}(k),
\end{eqnarray}
where~$ \underline{v} $ is a two component vector depending on the momentum variable $k \in [-\pi,\pi)$ (which becomes continuous in the thermodynamic limit $n\to\infty$). The eigenvalue problem reduces to the~following $2\times 2$ matrix problem
\begin{eqnarray}
  A\,\underline{v}(k)
  +\exp(\ii k)\, B\,\underline{v}(k)
  +\exp(-\ii k)\, C\,\underline{v}(k)
  -\lambda(k) \underline{v}(k)=0.
\end{eqnarray}
The eigenvalues and eigenvectors read
\begin{eqnarray}
  \eqalign{
	\lambda_{1,2}(k)=\frac{1}{2} \text{e}^{-\ii k}( \text{e}^{2 \ii k} \rho_1^2-2 \text{e}^{2 \ii k} \rho_1
      +2 \text{e}^{\ii k} \rho_2
	\rho_1+\text{e}^{2 \ii k}+ \rho_2^2-2 \rho_2+1\mp\\
  \phantom{\lambda_{1,2}(k)}\mp\left(\text{e}^{\ii k} \left( \rho_1-1\right)+ \rho_2-1\right) \delta)\\
	v_{1,2}(k)=
	\begin{bmatrix}
		\frac{\text{e}^{-\ii k}}{2\rho_2} \left(\text{e}^{\ii k} \left( \rho_1-1\right)- \rho_2+1\pm\delta\right),1
	\end{bmatrix},\\
	\delta=\sqrt{\text{e}^{2 \ii k}
		\left(1- \rho_1\right)^2+2 \text{e}^{\ii k} \left( \rho_2 \rho_1+ \rho_1+ \rho_2-1\right)
      +\left(1-\rho_2\right)^2}.
    }
\end{eqnarray}
In order to obtain the~full time evolution we express a~part of the~initial
state~$\mu_{2l-1} \hat{e}_{2l-1}+\mu_{2l}\hat{e}_{2l}$
in terms of the~eigenvectors,
\begin{eqnarray}
  \frac{\mu_{2l-1}}{\kappa_l}\hat{e}_{2l-1}+
  \frac{\mu_{2l}}{\kappa_l}\hat{e}_{2l}
  \!=\!\! \int_{-\pi}^{\pi}\!\!\!\!\dd k
 \bigoplus_r
 \e^{\ii k(r-l)}\big(\alpha_1(k)\ \underline{v}_{1}(k)+\alpha_2(k)\ \underline{v}_2(k)\big),\nonumber\\
 \kappa_l=\mu_{2l-1}+\mu_{2l}.
\end{eqnarray}
From this expression we can calculate the~constants~$\alpha_{1,2}(k)$,
yielding a~complete time-dependent profile~$f(x,t)$~\eqref{eq:chargeProfileDef}
in an~integral form,
\begin{equation}
  f(x,t)=\!\!\sum_{y=x-t}^{x+t}\kappa_y\int_{-\pi}^{\pi}\!\!\dd k\, e^{\ii k (x-y)}
  \bigg(
    \left(\lambda_1(k)\right)^t \tilde{\alpha}_1(k)
    +\left(\lambda_2(k)\right)^t \tilde{\alpha}_2(k)
  \bigg) ,
\end{equation}
where $ \tilde{\alpha}_{1,2}(k)=\alpha_{1,2}(k) \left([1,1]\cdot\underline{v}_{1,2}(k)\right)$.

Let us now focus on the~asymptotic shape of the~profile,~$t\to\infty$. In this limit we can
consider only the~contribution of the~leading eigenvalue,~$\lambda_1(k)$, since~$|\lambda_2(k)|<1$.
Furthermore, since~$\lambda_1(k)$ is an~analytic function of~$k$ in the~vicinity of~$k=0$
and~$\lambda_1(0)=1$, $|\lambda_1(k\neq0)|<1$, we should take into account only
the contributions at~$k\approx0$. In this region
the~leading eigenvalue can be approximated by
\begin{eqnarray}
  -\log \lambda_1(k)\approx i k \gamma_1 +\gamma_2 k^2,\nonumber\\
  \gamma_1=\frac{\rho_1-\rho_2}{\rho_1+\rho_2},
  \qquad \gamma_2=\frac{\rho_1 \rho_2
  \left(2-\rho_1-\rho_2\right)}{\left(\rho_1+\rho_2\right)^3},
\end{eqnarray}
which implies
\begin{eqnarray}
  \lambda_1(k)\approx\exp\left(-\ii \gamma_1 k -\gamma_2 k^2\right)
\end{eqnarray}

In the~long time limit, the steady states can form on different
space/time scales around the junction, depending on the~type of the~transport.
In particular, if the transport is ballistic, the~steady states arises
on the~light rays, $v=x/t$. However, in the case of diffusive transport
the~dynamics is more localized, therefore one should consider
the~steady state formation along the~space-time coordinates $u=x/\sqrt{t}$.

In general the~transport in our model is ballistic, therefore the~steady
state profile depends on the ballistic coordinate,
\begin{eqnarray}
  f(v)=\lim_{t\to\infty} f\left(\frac{n}{4}+v\, t,t\right).
\end{eqnarray}
The~only non-vanishing contribution steams from~$y$ in
the~vicinity of the~ballistic 
line,
\begin{eqnarray}
  y=n/4+(v-\gamma_1)t+u \sqrt{t}. 
\end{eqnarray}
If $u$ is kept finite as~$ t\to\infty$
and~$v-\gamma_1\neq 0$, one obtains
\begin{eqnarray}\fl
  f(v)=\big(\kappa^R H(v-\gamma_1)+\kappa^L
  H(\gamma_1-v)\big)
  \lim_{U\to\infty}\sum_{u=-U}^U \lim_{t\to\infty}
  \int_{-\pi}^{\pi}\dd k\,
  \mathrm{e}^{\ii k u\sqrt{t}-\gamma_2 k^2 t}
  \tilde{\alpha}_1(k),
\end{eqnarray}
where we introduced a~simpler notation~$\kappa^{L/R}=\mu_1^{L/R}+\mu_2^{L/R}$.
Introducing a~new variable~$ h=k\sqrt{t}$ and taking the~limit~$t\to\infty$,
the~integration is reduced to the~Gaussian integral. Noticing
that~$\tilde{\alpha}_1(0)=\frac{1}{2\pi}$,
we obtain the~following expression,
\begin{eqnarray}
  f(v)=\big(\kappa^{R} H(v-\gamma_1)+\kappa^L
  H(\gamma_1-v)\big)\lim_{U\to\infty}\sum_{u=-U}^U \frac{1}{2 \pi}
  \frac{\sqrt{\pi} \e^{-\frac{u^2}{4 \gamma_2}}}{\sqrt{t \gamma _2}}.
\end{eqnarray}
Approximating the~sum by an~integral and evaluating it
yields the~following result,
\begin{eqnarray}
  \label{ff}
  f(v)=\kappa^R H(v-\gamma_1)+\kappa^L H(\gamma_1-v).
\end{eqnarray}
The~asymptotic charge profile on ballistic time scales, $f(v)$,
is a~step function that
moves with the~velocity~$\gamma_1$.

Let us now consider the~diffusive region around the~ballistic
front,~$v-\gamma_1=\frac{u}{\sqrt{t}}$,
\begin{eqnarray}
  \tilde{f}(u)=\lim_{t\to\infty}f(\gamma_1+
  \tfrac{u}{\sqrt{t}})=\sum_{y=-n/4}^0 \kappa^L \int_{-\pi}^{\pi}
  \dd k\,\e^{\ii k u \sqrt{t}-\ii k y-\gamma_2 k^2 t}
  \tilde{\alpha}_1(k)
  +\nonumber\\
  +\sum_{y=1}^{n/4} \kappa^R \int_{-\pi}^{\pi}\dd k\,
  \e^{\ii k u \sqrt{t}-\ii k y-\gamma_2 k^2 t}
  \tilde{\alpha}_1(k).
\end{eqnarray}
Making similar simplifications as before, yields
\begin{eqnarray}
  \tilde{f}(u)=\frac{\kappa^L}{\sqrt{t}}\sum_{y=-\infty}^0 \frac{1}{2 \pi}
  \frac{\sqrt{\pi } e^{-\frac{\left(y/\sqrt{t}- u\right)^2}{4 \gamma_2
  }}}{\sqrt{\gamma_2}}+\frac{\kappa^R}{\sqrt{t}}\sum_{y=1}^\infty \frac{1}{2
  \pi} \frac{\sqrt{\pi } e^{-\frac{\left(y/\sqrt{t}- u\right)^2}{4 \gamma_2
  }}}{\sqrt{\gamma_2}},
\end{eqnarray}
and after the~identification of the~sums with integrals
we get the~final result
\begin{eqnarray}
  \label{difres}
  \tilde{f}(u)=\half(\kappa^L+\kappa^R)+\half (\kappa^R-\kappa^L)\, \mathrm{erf}\left(\frac{u}{2\sqrt{\gamma_2}}\right).
\end{eqnarray}
The~solution of the~diffusion equation,
$\frac{\partial}{\partial t} q(x,t)=\mathcal{D} \frac{\partial^2}{\partial x^2}
q(x,t)$,
with the~diffusion constant $ \mathcal{D} $ is
\begin{eqnarray}
  q(x,t)=\mathrm{erf}\left(\frac{x}{\sqrt{4 \mathcal{D} t}}\right),
\end{eqnarray}
therefore we can read out the~diffusion constant
from the~expression~\eqref{difres}. Since the~coordinate~$x$ corresponds
to two lattice sites, the~diffusion constant should be rescaled,
\begin{eqnarray}
  \mathcal{D}=4\gamma_2=
  \frac{4\rho_1\rho_2(2-\rho_1-\rho_2)}{\left(\rho_1+\rho_2\right)^3}.
\end{eqnarray}

\subsection{The hydrodynamical picture}
Here we derive the velocity of the front~$ \gamma_1 $ using the
arguments of generalized hydrodynamics. The basic idea is the following: considering two half chains prepared in
distinct stationary homogeneous states joined at the origin. The hydrodynamical approach assumes that the~non-equilibrium
steady state arises along the light-cone coordinates centred at the origin. The system is
assumed to reach generalized equilibrium~$ \rho_{GGE}(v) $ on a given 
light-ray~$ v=x/t $ \cite{bertini,PhysRevX.6.041065}. Note that this is not surprising, 
since in the~limit~$t\to\infty$,
the subsystem between the~light-rays~$ v\pm \varepsilon$ is infinitely large
and is expected to equilibrate. The remaining slow modes characterizing
the~NESS are the conserved charges and their currents.

We wish to compute
the profiles of the charge $ q $ and the corresponding current $ j $.
Their expectation values in the~GGE are
\begin{eqnarray}
  \label{exp1} \fl \langle q
  \rangle_{\tilde{\rho}_{GGE}(v)}=(\mu_2-\mu_1)c^-_1- (\mu_1+\mu_2)c^+_1+
  (\mu_1+\mu_2)c_0\\
  \label{exp2} \fl
  \langle j
  \rangle_{\tilde{\rho}_{GGE}(v)}=-2 (\mu_1+\mu_2)
  c^{-}_1+2\big(\mu_2(1-2\rho_1)-\mu_1 (1-2 \rho_2)\big)c_1^+ +\nonumber\\
  +2\big(\mu_1 (1- \rho_2)-\mu_2(1-\rho_1)\big) c_0,
\end{eqnarray}
where the constants $ c_0,\ c^+_1,\ c^-_1 $ are determined by fixing expectation values of
the~charges,
\begin{eqnarray}
  \langle s^{\pm}_1 \rangle_{\tilde{\rho}_{GGE}(v)}=0,\qquad
  \langle 1 \rangle_{\tilde{\rho}_{GGE}(v)}=1.
\end{eqnarray}
The only coefficient that depends on the light-ray is~$\mu$ defined
as~$\mu_1=\mu\, \rho_1,\ \mu_2=\mu\, \rho_2 $. 
Inserting expectation
values~\eqref{exp1},~\eqref{exp2} into the~continuity equation,
\begin{eqnarray}
  \label{cont}
  \partial_t\langle q \rangle_{\tilde{\rho}_{GGE}(v)}+\partial_x \langle j \rangle_{\tilde{\rho}_{GGE}(v)}=0,
\end{eqnarray}
yields the~following equation for the chemical potential~$ \mu $
\begin{eqnarray}
  (\rho_1+\rho_2)\,\partial_t \mu+2 (\rho_1-\rho_2)\,\partial_x \mu=0.
\end{eqnarray}
Taking into account the~initial condition we obtain the~solution
for~$ \mu $,
\begin{eqnarray}
  \left(\rho_1+\rho_2\right)\mu(x,t)=\kappa^R H(v-\gamma_1)+\kappa^L H(\gamma_1-v),
\end{eqnarray}
which is in perfect accordance with~\eqref{ff}.

\section{Dynamic~structure factor}\label{sec:StructureFactor}
Finally, we calculate the full spatio-temporal density-density correlation function $C_q(x,t)=\ave{q_0^t\,q_x}_p$. For simplicity, we restrict
the discussion to the~deterministic, translationally invariant case
(i.e.\ $\rho_1=\rho_2$ and~$\Gamma=0$). Similarly as before, only the observables
with at most one~local~$\oo{2}$ contribute to the overlap~$\ave{q_0^t\,q_x}$,
therefore the time evolution of~$q_0$ can be restricted to the infinite family
of subalgebras~$\mathcal{A}^{[z]}$,~$z\in\ZZ$,
\begin{eqnarray}
  \mathcal{A}^{[0]}
    =\lsp\{a^{\text{e}\,0}_x,
    a^{\text{o}\,0}_x;x\in\ZZ\},\qquad
    \mathcal{A}^{[z\neq 0]}=\lsp\{a^z_x;x\in\ZZ\},
\end{eqnarray}
with the basis~elements~$a^{z}_x$ defined as
\begin{eqnarray}
\label{bal}
  \eqalign{
    &a^{\text{e}\,0}_x = \oo{1 0}\emptysub{2x},\\
    &a^{2d+1}_x=\oo{0 1 \underbrace{0 \cdots 0}_{2d} 2}\emptysub{2x},\\
    &a^{2d+2}_x=\oo{1 \underbrace{0 \cdots 0}_{2d+1} 2}\emptysub{2x},
  }\qquad
  \eqalign{
    &a^{\text{o}\,0}_x = \oo{0 1}\emptysub{2x},\\
    &a^{-(2d+1)}_x=\oo{0 2 \underbrace{0 \cdots 0}_{2d} 1}\emptysub{2x-(2d+2)},\\
    &a^{-(2d+2)}=\oo{2 \underbrace{0 \cdots 0}_{2d+1} 1}\emptysub{2x-(2d+2)}, 
  }
\end{eqnarray}
The~reduced full time step propagator~$\mathcal{U}$ has the~following matrix elements,
\begin{eqnarray}\label{eq:structureFactorReducedPropagator}
  \mathcal{U}^{d^{\prime}\,d}_{x^{\prime}\,x} = \big\langle
  \big(a^{d^{\prime}}_{x^{\prime}}\big)^{\prime} U a^{d}_x
  \big\rangle.
\end{eqnarray}
The dual vectors $(a^d_x)'$ are obtained from expressions (\ref{bal}) simply by replacing the canonical basis vectors by the corresponding dual (primed) vectors.
In this basis, the dynamical structure factor $C_q(x,t)$ can be expressed as
\begin{eqnarray}\label{eq:structurefactororiginalexpression}
  C_q(2x,t)=
    \langle
    \left(a_{x}^{\mathrm{e}\,0}+a_{x}^{\mathrm{o}\,0} \right)
    U^{t}
    \left(a_0^{\mathrm{e}\,0}+ a_0^{\mathrm{o}\,0}\right)
    \rangle_p \equiv O_x\,\mathcal{U}^tQ_0,
\end{eqnarray}
where we introduced the~initial vector~$Q_0$ corresponding to the charge at  the origin~$a_0^{\mathrm{e}\,0}+ a_0^{\mathrm{o}\,0}$
\begin{eqnarray}
\left[Q_0\right]_y^d =\delta_{d,0}\delta_{y,0},
\end{eqnarray}
and the vector~$O_x$ that encodes the~overlaps
between the charge at the position $ x $, $a_x^{\mathrm{e}\,0}+ a_x^{\mathrm{o}\,0}$ and the basis elements
\eqref{bal}
\begin{eqnarray}
  \left[O_x\right]_y^d =
    \langle 
    a_x^{\mathrm{e}\,0} a_y^{d}
    \rangle_p+
    \langle 
    a_x^{\mathrm{o}\,0} a_y^{d}
    \rangle_p.
\end{eqnarray}
The~submatrices~$\mathcal{U}^{d^{\prime}\, d}$, with the~elements
$[ \mathcal{U}^{d^{\prime}\, d}]_{x^{\prime}\, x} = \mathcal{U}^{d^{\prime}\, d}_{x^{\prime}\,x}$,
are mostly zero, due to the following property of the reduced time propagator
\begin{eqnarray}
  \eqalign{
    \mathcal{U}:\ & \\
    &\mathcal{A}^{[0]} \to \mathcal{A}^{[0]}, \mathcal{A}^{[\pm 1]}, \mathcal{A}^{[\pm 2]}, \mathcal{A}^{[\pm 3]},\\
    &\mathcal{A}^{[\pm(2d-1)]} \to
    \mathcal{A}^{[\pm 2d]}, \mathcal{A}^{[\pm (2d+1)]}, \mathcal{A}^{[\pm(2d+2)]}, \mathcal{A}^{[\pm(2d+3)]},\\
    &\mathcal{A}^{[\pm 2d]} \to
    \mathcal{A}^{[\pm 2d]}, \mathcal{A}^{[\pm (2d+1)]}, \mathcal{A}^{[\pm(2d+2)]}, \mathcal{A}^{[\pm(2d+3)]}.
  }
\end{eqnarray}
Furthermore, the submatrices~$\mathcal{U}^{d^{\prime}\, d}$ have block three-diagonal structure,
with the block dimensions~$1\times 1$, $1\times 2$ and~$2\times 2$
(for explicit expression see~\ref{app:structureFactor}), therefore it is possible to
use a~similar approach as in section~\ref{sec:inhomogeneousQuench}.
Introducing the~Fourier basis,
\begin{eqnarray}\label{eq:fourierBasis}
  \eqalign{
    e(k)=\sum_x \mathrm{e}^{\ii k x}a_x^{\text{e}\,0},\qquad
    o(k)=\sum_x \mathrm{e}^{\ii k x}a_x^{\text{o}\,0},\\
    g^{[d]}(k)=\sum_x \mathrm{e}^{\ii k x} a_x^{d},
  }
\end{eqnarray}
each of the~infinite submatrices~$\mathcal{U}^{d^{\prime}\,d}$ is reduced to a~finite matrix, which depends on the~parameter~$k$ (see~\ref{app:structureFactor} for the~details). 
In the~basis
\begin{eqnarray}
  \begin{bmatrix}
    \ldots & g^{[-2]} (k) & g^{[-1]} (k) & e(k) & o(k) & g^{[1]}(k) & g^{[2]}(k) & \ldots
  \end{bmatrix},
\end{eqnarray}
the time propagator takes the following form,
\begin{eqnarray} \label{eq:InfFourierMatrix}
  \widetilde{\mathcal{U}}=
  \begin{tikzpicture}[baseline=(current bounding box.center),
      every right delimiter/.style={xshift=-2mm},
    every left delimiter/.style={xshift=2mm}]
    \matrix (M)[matrix of math nodes, nodes in empty cells,
      left delimiter={[}, right delimiter={]}, every node/.style={minimum width={5mm}, minimum height={5mm}},
    ]{&&&&&&&&&&&\\
      &&&&&&&&&&&\\
      &&&&&&&&&&&\\
      &&&&&&&&&&&\\
      &&&&&&&&&&&\\
      &&&&&&&&&&&\\
      &&&&&&&&&&&\\
      &&&&&&&&&&&\\
      &&&&&&&&&&&\\
      &&&&&&&&&&&\\
      &&&&&&&&&&&\\
      &&&&&&&&&&&\\
      &&&&&&&&&&&\\
      &&&&&&&&&&&\\
      &&&&&&&&&&&\\
      &&&&&&&&&&&\\
      &&&&&&&&&&&\\
    &&&&&&&&&&&\\};
    \begin{scope}[on background layer]
      \draw[draw opacity=0.5,color=green] (M-1-1.north east)   -- (M-5-1.south east);
      \draw[draw opacity=0.5,color=green] (M-2-2.north east)   -- (M-5-2.south east);
      \draw[draw opacity=0.5,color=green] (M-2-3.north east)   -- (M-7-3.south east);
      \draw[draw opacity=0.5,color=green] (M-4-4.north east)   -- (M-7-4.south east);
      \draw[draw opacity=0.5,color=green] (M-4-5.north east)   -- (M-5-5.south east);
      \draw[draw opacity=0.4,color=red]   (M-6-5.north east)   -- (M-13-5.south east);
      \draw[draw opacity=0.4,color=red]   (M-6-6.north east)   -- (M-13-6.south east);
      \draw[draw opacity=0.4,color=red]   (M-6-7.north east)   -- (M-13-7.south east);
      \draw[draw opacity=0.5,color=green] (M-14-7.north east)  -- (M-15-7.south east);
      \draw[draw opacity=0.5,color=green] (M-12-8.north east)  -- (M-15-8.south east);
      \draw[draw opacity=0.5,color=green] (M-12-9.north east)  -- (M-17-9.south east);
      \draw[draw opacity=0.5,color=green] (M-14-10.north east) -- (M-17-10.south east);
      \draw[draw opacity=0.5,color=green] (M-14-11.north east) -- (M-18-11.south east);

      \foreach \x in {2,...,4}
      \draw[draw opacity=0.5,color=green] (M-\x-1.north west)   -- (M-\x-1.north east);
      \foreach \x in {2,...,6}
      \draw[draw opacity=0.5,color=green] (M-\x-2.north west)   -- (M-\x-3.north east);
      \foreach \x in {4,...,8}
      \draw[draw opacity=0.5,color=green] (M-\x-4.north west)   -- (M-\x-5.north east);
      \foreach \x in {6,...,14}
      \draw[draw opacity=0.4,color=red]   (M-\x-6.north west)   -- (M-\x-7.north east);
      \foreach \x in {12,...,16}
      \draw[draw opacity=0.5,color=green] (M-\x-8.north west)   -- (M-\x-9.north east);
      \foreach \x in {14,...,18}
      \draw[draw opacity=0.5,color=green] (M-\x-10.north west)   -- (M-\x-11.north east);
      \foreach \x in {16,...,18}
      \draw[draw opacity=0.5,color=green] (M-\x-12.north west)   -- (M-\x-12.north east);

      \draw[text opacity=1,draw opacity=0,fill=green, fill opacity=0.2] (M-1-1.north west) rectangle (M-3-1.south east)
      node[midway]{$\ddots$};
      \draw[text opacity=1,draw opacity=0,fill=green, fill opacity=0.2] (M-16-12.north west) rectangle (M-18-12.south east)
      node[midway]{$\ddots$};
      \draw[text opacity=1,draw opacity=0,fill=green, fill opacity=0.2] (M-2-2.north west) rectangle (M-5-3.south east)
      node[midway]{$B$};
      \draw[text opacity=1,draw opacity=0,fill=green, fill opacity=0.2] (M-4-4.north west) rectangle (M-7-5.south east)
      node[midway]{$B$};
      \draw[text opacity=1,draw opacity=0,fill=green, fill opacity=0.2] (M-12-8.north west) rectangle (M-15-9.south east)
      node[midway]{$B$};
      \draw[text opacity=1,draw opacity=0,fill=green, fill opacity=0.2] (M-14-10.north west) rectangle (M-17-11.south east)
      node[midway]{$B$};
      \draw[text opacity=1,draw opacity=0,fill=red, fill opacity=0.2] (M-6-6.north west) rectangle (M-13-7.south east)
      node[midway]{$A$};
    \end{scope}
  \end{tikzpicture},
\end{eqnarray}
with the matrices~$A$ and~$B$ given by
\begin{eqnarray} \label{eq:InfFourierMatrixBlocks}
  \eqalign{
    &A = \begin{bmatrix}
      \mathrm{e}^{-\ii k}(1-\rho)^2 & -\mathrm{e}^{-\ii k}(1-\rho)^2 \\
      \rho(1-\rho) & -\rho(1-\rho) \\
      \mathrm{e}^{-\ii k}(1-\rho)^2-\rho(1-\rho) & -(1-\rho)^2+\mathrm{e}^{-\ii k}\rho(1-\rho)\\
      \rho^2+(1-\rho)^2\mathrm{e}^{-\ii k} & \rho(1-\rho)(1+\mathrm{e}^{-\ii k}) \\
      \rho(1-\rho)(1+\mathrm{e}^{\ii k}) & \rho^2+(1-\rho)^2 \mathrm{e}^{\ii k}\\
      \mathrm{e}^{\ii k}\rho (1-\rho)-(1-\rho)^2 & -\rho (1-\rho)+\mathrm{e}^{\ii k}(1-\rho)^2\\
      -\rho(1-\rho) & \rho(1-\rho)\\
      -\mathrm{e}^{\ii k}(1-\rho)^2 & \mathrm{e}^{\ii k}(1-\rho)^2\\
    \end{bmatrix},\\
    &B = \begin{bmatrix}
      \mathrm{e}^{-\ii k}\rho(1-\rho) & \mathrm{e}^{-\ii k}(1-\rho)^2 \\
      \rho^2 & \rho(1-\rho) \\
      \rho(1-\rho) & \rho^2\\
      \mathrm{e}^{\ii k}(1-\rho)^2 & \mathrm{e}^{\ii k}\rho (1-\rho)\\
    \end{bmatrix},
  }
\end{eqnarray}
while the initial vector~$\mathcal{Q}_0$ and the overlap vector~$\mathcal{O}_x$ are expressed as
\begin{eqnarray}
  \eqalign{
    \mathcal{O}_x&=\mathrm{e}^{\ii k x}\left(\rho\,\mathcal{O}^{\text{(D)}}-\mu^2\mathcal{O}^{\text{(B)}}\right),\\
    \mathcal{O}^{\text{(D)}}&=
    \begin{bmatrix}\ldots & 
      0&
      0&
      0&
      1&
      1&
      0&
      0&
      0&
      \ldots
    \end{bmatrix},\\
    \mathcal{O}^{\text{(B)}}&=
    \begin{bmatrix}\ldots &
      \mathrm{e}^{2 \ii k}&
      \mathrm{e}^{2 \ii k}&
      \mathrm{e}^{\ii k}&
      \mathrm{e}^{\ii k}&
      1&
      1&
      \mathrm{e}^{-\ii k}&
      \mathrm{e}^{-\ii k}&
      \mathrm{e}^{-2 \ii k}&
      \mathrm{e}^{-2 \ii k}&
      \ldots
    \end{bmatrix},\\
    \mathcal{Q}_0 &= \begin{bmatrix}\ldots & 0 & 0 & 0 & 1 & 1 & 0 & 0 & 0 &\ldots\end{bmatrix}.
  }
\end{eqnarray}
In the Fourier basis~\eqref{eq:structurefactororiginalexpression}, the dynamical structure factor corresponds to
\begin{eqnarray}
  C_q(2 x, t)=\int_{-\pi}^\pi \dd k\,\mathcal{O}_x\,\widetilde{\mathcal{U}}^t \mathcal{Q}_0.
\end{eqnarray}
The calculation can be split into a~$\rho$-dependent diffusive part and a~ballistic contribution proportional
to $\mu^2$,
\begin{eqnarray}
  C_q(x,t)=\rho\,C_q^\mathrm{(D)}(x,t)-\mu^2 C_q^\mathrm{(B)}(x,t).
\end{eqnarray}

The diffusive contribution can be evaluated rather easily,
since~$\mathcal{O}^\mathrm{(D)}=\mathcal{Q}_0$,
\begin{eqnarray}\label{eq:corDiffusiveContribution}
  C_q^\mathrm{(D)}(2 x, t)=\int_{-\pi}^\pi \dd k \,
  \mathrm{e}^{\mathrm{i} k x}
  \begin{bmatrix}
    1 & 1
  \end{bmatrix}
  \widetilde{\mathcal{U}}^{\mathrm{(D)}t}
  \begin{bmatrix}
    1 \\ 1
  \end{bmatrix},
\end{eqnarray}
where the~matrix $\widetilde{\mathcal{U}}^{\mathrm{(D)}}$ corresponds to the central $2\times 2$ block of
the reduced propagator $ \mathcal{U} $~\eqref{eq:InfFourierMatrix},
\begin{eqnarray}
  \widetilde{\mathcal{U}}^{\mathrm{(D)}}=
  \begin{bmatrix}
    \rho^2+(1-\rho)^2\mathrm{e}^{-\ii k} & \rho(1-\rho)(1+\mathrm{e}^{-\ii k}) \\
    \rho(1-\rho)(1+\mathrm{e}^{\ii k}) & \rho^2+(1-\rho)^2 \mathrm{e}^{\ii k}\\
  \end{bmatrix}.
\end{eqnarray}
In the large time limit, the contribution corresponds to the normal distribution,
\begin{eqnarray}
  C_q^{\text{(D)}}(x,t) = \frac{1}{\sqrt{t\pi\mathcal{D}}}\mathrm{e}^{-\frac{x^2}{4\mathcal{D}t}},
\end{eqnarray}
with $\mathcal{D}=\rho^{-1}-1$.

\begin{figure}
  \centering
  \includegraphics{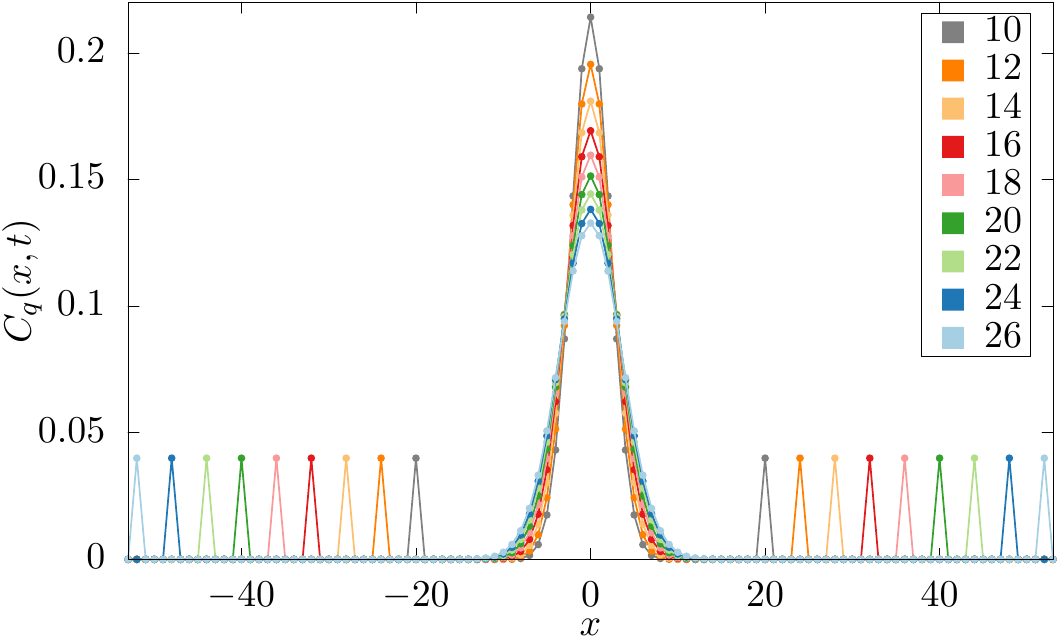}
  \caption{\label{fig:spatiotemporal} An example of the asymptotic correlation profile~$C_q(x,t)$
    at different times (denoted by different colors), as described by~\eqref{eq:wholespatiotemporal},
  with~$\rho=0.8$ and~$\mu=0.4$.}
\end{figure}

In order to obtain the ballistic contribution, we first observe that
for any constants $c_{\text{e/o}}\in \mathbb{C}$, the following holds
\begin{eqnarray}\fl
  \begin{bmatrix}\ldots &
    \mathrm{e}^{\ii(t+2) k }&
    \mathrm{e}^{\ii (t+1) k}&
    \mathrm{e}^{\ii (t+1) k}&
    c_{\text{e}} & c_{\text{o}} &
    \mathrm{e}^{-\ii (t+1) k}&
    \mathrm{e}^{-\ii (t+1) k}&
    \ldots
  \end{bmatrix} \widetilde{\mathcal{U}}=\nonumber\\\fl
  =\begin{bmatrix}\ldots &
    \mathrm{e}^{\ii(t+3)k}&
    \mathrm{e}^{\ii(t+2)k}&
    \mathrm{e}^{\ii(t+2)k}&
  c_{\text{e}}^{\prime} & c_{\text{o}}^{\prime} &
    \mathrm{e}^{-\ii(t+2)k}&
    \mathrm{e}^{-\ii(t+2)k}&
    \ldots
  \end{bmatrix},
\end{eqnarray}
with appropriate values of $c_{\text{e/o}}^{\prime}\in\mathbb{C}$.
This implies that the ballistic part $C_q^{\text{(B)}}$ can be encoded in terms of the reduced vectors
\begin{eqnarray}
\eqalign{
	\tilde{\mathcal{O}}^\mathrm{(B)}&=\begin{bmatrix}\mathrm{e}^{2 \ii k}& \mathrm{e}^{\ii k} & \mathrm{e}^{\ii k} & 1 & 1 &
	\mathrm{e}^{-\ii k}
	& \mathrm{e}^{-\ii k}
	& \mathrm{e}^{-2 \ii k}\end{bmatrix},\\
	\tilde{\mathcal{Q}}_0^T&=\begin{bmatrix} 0 & 0 & 0 & 1 & 1 & 0 & 0 & 0\end{bmatrix},
}
\end{eqnarray}
and the matrix
\begin{eqnarray}\fl
\scalemath{0.9}{
	\widetilde{\mathcal{U}}^{\mathrm{(B)}}\!\!=\!\!
	\begin{bmatrix}
	\mathrm{e}^{\ii k}& & & \mathrm{e}^{-\ii k}(1-\rho)^2 & -\mathrm{e}^{-\ii k}(1-\rho)^2 \\
	&\mathrm{e}^{\ii k} & & \rho(1-\rho) & -\rho(1-\rho) \\
	& & \mathrm{e}^{\ii k}& \mathrm{e}^{-\ii k}(1-\rho)^2-\rho(1-\rho) & -(1-\rho)^2+\mathrm{e}^{-\ii k}\rho(1-\rho)\\
	& & & \rho^2+(1-\rho)^2\mathrm{e}^{-\ii k} & \rho(1-\rho)(1+\mathrm{e}^{-\ii k}) \\
	& & & \rho(1-\rho)(1+\mathrm{e}^{\ii k}) & \rho^2+(1-\rho)^2 \mathrm{e}^{\ii k}\\
	& & & \mathrm{e}^{\ii k}\rho (1-\rho)-(1-\rho)^2 & -\rho (1-\rho)+\mathrm{e}^{\ii k}(1-\rho)^2 & \mathrm{e}^{-\ii k}\\
	& & & -\rho(1-\rho) & \rho(1-\rho) & &  \mathrm{e}^{-\ii k} \\
	& & & -\mathrm{e}^{\ii k}(1-\rho)^2 & \mathrm{e}^{\ii k}(1-\rho)^2  & & &  \mathrm{e}^{-\ii k}\\
	\end{bmatrix}
},
\end{eqnarray}
as
\begin{eqnarray}
  C_q^{\mathrm{(B)}}(2x,t)=\int_{-\pi}^{\pi} \dd k \, \mathrm{e}^{\ii k x}
  \tilde{\mathcal{O}}^\mathrm{(B)}\widetilde{\mathcal{U}}^{\mathrm{(B)}\,t} \tilde{\mathcal{Q}}_0.
\end{eqnarray}
Diagonalizing the transposed matrix~$\widetilde{\mathcal{U}}^{\mathrm{(B)}T}$ and calculating the relevant overlaps, we obtain
\begin{eqnarray}
  \eqalign{
  C_q^{\mathrm{(B)}}(2x,t)&=
  -\frac{1-\rho}{2\pi\rho}\!\!\int_{-\pi}^{\pi}\!\!\!\dd k \left(\e^{-\ii k (x-t)}+\e^{\ii k (x+t)}\right)+
  \\&+
  \!\!\int^\pi_{-\pi}\!\!\!\dd k\,\e^{\ii k x}\left(
    \widetilde{\alpha}_1(k)\lambda_1(k)^t 
    +\widetilde{\alpha}_2(k)\lambda_2(k)^t 
  \right),
}
\end{eqnarray}
with
\begin{eqnarray}\fl
  \widetilde{\alpha}_{1,2}(k)=\frac{1}{2\pi\rho} \pm \frac{1+\mathrm{e}^{\ii k}}{2\pi\sqrt{2\mathrm{e}^{\ii k}
  \left((1-\rho)^2\cos k +\rho(2+\rho)-1\right)}},\\
  \fl\nonumber
  \lambda_{1,2}(k) = \rho^2+(1-\rho)^2\cos k \pm\\
  \pm(1-\rho)(1+\mathrm{e}^{-\ii k}){\sqrt{\frac{1}{2}\mathrm{e}^{\ii k}
  \Big((1-\rho)^2\cos k +\rho(2+\rho)-1\Big)}}.
\end{eqnarray}
In the large time limit, the~first integral reduces
to~$\delta_{x,t}+\delta_{x,-t}$, while the second one correspond to the normal
distribution~\eqref{eq:corDiffusiveContribution}.
Taking into account both of the~contributions, we finally 
obtain the~asymptotic shape of the structure factor,
\begin{eqnarray}\label{eq:wholespatiotemporal}
  C_q(x,t) = \rho\left(1-\frac{\mu^2}{\rho^2}\right)\frac{1}{\sqrt{t\pi\mathcal{D}}}\,\e^{-\frac{x^2}{4\mathcal{D}t}}
  +\mu^2\mathcal{D}\left(\delta_{x/2,t}+\delta_{-x/2,t}\right),
\end{eqnarray}
with~$\mathcal{D}=\rho^{-1}-1$.
The~profile consists of the~central peak that
spreads diffusively, and two spikes that propagate ballistically, see~Fig.~\ref{fig:spatiotemporal}.

\section{Conclusion}
In the paper we explored the transport properties of a~simple reversible cellular
automaton modelling a hard-point interacting gas of charged particles. We constructed the set of conserved quantities and
the~corresponding set of stationary states described by the~Generalized Gibbs Ensemble.
By obtaining the~explicit expressions for the time dependence of autocorrelation
functions we were able to calculate the Drude weight and the diffusion constant.
The Drude weight was shown to match the Mazur lower bound perfectly.  We
also analytically solved the inhomogeneous initial state problem, which
corresponds to the step function charge density profile moving with a~constant
velocity and the~diffusive corrections at large times. The velocity of
the step function profile was shown to match the velocity obtained from the
hydrodynamic consideration. Furthermore, we calculated the~structure
factor exactly. The asymptotic spatio-temporal correlation profile consists
of two ballistically moving~$\delta$-spikes and a~diffusively broadening central
peak.

Our model can be used as a rigorous benchmark of the physical properties of locally interacting lattice systems with conservation laws,
showing agreement with the proposed effective descriptions such as hydrodynamics or the Mazur inequality, which are typically
used without rigorous justification. The main question that remains is whether a similar description of complete
time evolution can be obtained for more complicated interacting classical or quantum
systems, say for a typical integrable systems.

\section*{Acknowledgements}
We thank H Spohn for stimulating discussions and V Popkov and M Vanicat for ongoing collaboration on related problems.
The work has been supported by Advanced Grant 694544 -- OMNES of European
Research Council (ERC), and by Research program P1-0044 of 
Slovenian Research Agency (ARRS).

\section*{References}
\bibliographystyle{iopart-num}
\bibliography{dif_cl}
\appendix
\section{The~linear response}\label{app:linearResponse}
Here, we clarify how to derive the linear response coefficients. Note that the
approach used in the supplementary material of~\cite{PhysRevLett.119.110603}
yields correct expressions in the case of deterministic reversible dynamics.
However, when considering the stochastic generalization, the time propagator
is no longer a automorphism, therefore one should be more careful.

The continuity equation relates the time derivative of the density with the
spatial derivative of the current, and should hold for any initial state~$p$,
\begin{eqnarray}\label{eq:continuityState}
  \langle q_x\, p^{t+1} \rangle-\langle q_x\, p^{t} \rangle +\half
  \left(\langle j_{x+1}\,p^{t+\half} \rangle -\langle j_{x-1}\,p^{t+\half}
  \rangle\right)=0,
\end{eqnarray}
where $p^{t+{\frac{1}{2}}}$ denotes the state at the intermediate
half-time slice $t+\frac{1}{2}$,
i.e.\ $p^{t+\frac{1}{2}}=U^{\text{o}}p^t$.
On the level of observables, the continuity equation~\eqref{eq:continuityState} reads
\begin{eqnarray}
  U^\text{o} U^\text{e}q_x-q_x+\half \left(U^\text{o} j_{x+1}-U^\text{o} j_{x-1}\right)=0.
\end{eqnarray}
If $ x\in 2 \ZZ $ this relation reduces to
\begin{eqnarray}
  \label{sodi}
  U^\text{o} q_x-q_x+\half \left(U^\text{o} j_{x+1}-U^\text{o} j_{x-1}\right)=0;\quad x\in 2 \ZZ,
\end{eqnarray}
due to the invariance of $q_x$ under the $U^\text{e}$; $U^{\text{e}} q_x=q_x$. If, however,
$x$ is odd, the inverse of the propagator $U^{\text{o}}$ conserves the charge,
$ (U^\text{o})^{-1}q_x=q_x $, which yields the following relation
\begin{eqnarray}
  \label{lihi}
  U^\text{e} q_x-q_x+\half (j_{x+1}-j_{x-1})=0;\quad x\in 2\ZZ+1.
\end{eqnarray}
Since the two relations \eqref{sodi} and \eqref{lihi} have a~different form, the
current has a staggered structure. If the time evolution is deterministic,
the local propagator is a permutation
$(U^\text{o})^{-1}=U^\text{o} $, which implies the following relation
between the currents on odd and even sites,
\begin{eqnarray}\label{eq:appAOddEvenDeterministic}
  j_{x+1}=(-1)\,\eta_1( j_{x}).
\end{eqnarray}
In general, however, the connection between the odd and the even currents can
be deduced directly from the relations \eqref{lihi} and \eqref{sodi}, and reads
\begin{eqnarray}
  U^\text{o}j_{2 x+1}=\eta_1(j_{2 x}).
\end{eqnarray}

The linear response quench setup is consistent with the one considered in
\cite{PhysRevLett.119.110603}. The system in the initial stationary state $ p $
is kicked out of equilibrium at time $ t=0 $, by a weak constant external
field~$K(h)$ -- a constant force $h$:
\begin{eqnarray}
  K(h)=1+h\sum_{x=-n}^n x q_x. 
\end{eqnarray}
We note that for convenience we (only here) label our spatial lattice from
$x=-n$ to $x=n$, thus consisting of $2n+1$ sites and considering open boundary
conditions.  At time~$t$, the current induced by the force takes the following
form
\begin{eqnarray}
  j_{LR}(t)=\half\left(\langle (j_0^{\frac{1}{2}}+j_1^{\frac{1}{2}})
    \left(U^{\text{o}}U^{t}\right) K(h) \rangle_p -\langle(
    j_0^{\frac{1}{2}}+j_1^{\frac{1}{2}})U^{\text{o}}K(h)
  \rangle_p\right).
\end{eqnarray}
Due to the staggered structure, the current is averaged over two lattice sites
and because of the discreteness of time, it is additionally shifted for half of
the time step (see~\eqref{eq:continuityState}). To make the notation more compact
we introduced the following convention
\begin{eqnarray}
  j^{t+\frac{1}{2}}_x=U^{\text{o}}j_x^t=U^{\text{o}}U^t j_x.
\end{eqnarray}
Note that we subtracted the
initial value of the current, since we are not interested in the current
already present in the initial state, but rather in the current that is induced
on top of the initial state current due to the perturbation.

Writing out the propagator explicitly,
\begin{eqnarray}
  j_{LR}(t)=\half h\sum_{|x|\leq n} x \langle
  (j_0^{\frac{1}{2}}+j_1^{\frac{1}{2}})U^{\text{o}}
  (U^{\text{e}}U^{\text{o}})^t q_x\rangle_p,
\end{eqnarray}
and dividing the expression into even and odd contributions yields
\begin{eqnarray}\fl
  \label{eq2}
  \frac{2 j_{LR}(t)}{h}=
  \smashoperator{\sum_{x\in 2\ZZ;\,|x|\leq n}}
  x \langle
  (j_0^{\frac{1}{2}}+j_1^{\frac{1}{2}})U^{\text{o}}(U^\text{e} U^\text{o})^t q_x \rangle_p +
  \smashoperator{\sum_{x\in 2\ZZ+1;\,|x|\leq n}}
  x \langle (j_0^{\frac{1}{2}}+j_1^{\frac{1}{2}})U^{\text{o}} (U^\text{e} U^\text{o})^t q_x \rangle_p.
\end{eqnarray}
Taking into account the continuity equation, the expression \eqref{eq2} can be recast into
\begin{eqnarray}\fl
  \frac{2 j_{LR}(t)}{h}=
  \smashoperator{\sum_{\substack{{x\in 2\ZZ;\,|x|\leq n}\\ 1\le t \le \tau}}}
  \langle (j_0^{\frac{1}{2}}+j_1^{\frac{1}{2}})U^{\text{o}}(U^\text{e} U^\text{o})^\tau j_x \rangle_p
  +\smashoperator{\sum_{\substack{{x\in 2\ZZ +1;\,|x|\leq n}\\ 1\le t \le \tau}}}
\langle (j_0^{\frac{1}{2}}+j_1^{\frac{1}{2}})U^{\text{o}}(U^\text{e} U^\text{o})^{(\tau-1)} j_x \rangle_p\Bigg),
\end{eqnarray}
therefore the current at time $t$ is equal to
\begin{eqnarray}
  \label{tk2}
  \frac{j_{LR}(t)}{h}=\half \left(\smashoperator[r]{\sum_{x\in 2\ZZ+1;\,|x|\leq n}}
    \langle (j_0^{\frac{1}{2}}+j_1^{\frac{1}{2}}) j_x^{\frac{1}{2}} \rangle_p
  +\sum_{\tau=1}^{t} \sum_{|x|\leq n} \langle (j_0^{\frac{1}{2}}+j_1^{\frac{1}{2}}) j^{t+\frac{1}{2}}_x\rangle_p\right).
\end{eqnarray}
Note that in the deterministic case, we can take into account the simpler relation between
the currents on odd and even sites~\eqref{eq:appAOddEvenDeterministic} and
the homomorphism property of $U^{\text{o}}$ to reduce the expression to
\begin{eqnarray}
  \frac{j_{LR}(t)}{h}=\half \left(\sum_{|x|\leq n}  \half\langle (j_0+j_1) j_x
  \rangle_p+\sum_{\tau=1}^{t} \sum_{|x|\leq n}
\langle (j_0+j_1) j^t_x\rangle_p\right).
\end{eqnarray}
The asymptotic value of the current corresponds to the conductivity
\begin{eqnarray}
  \sigma=\lim_{t\to \infty} \lim_{n\to\infty}\lim_{h\to 0} \frac{j_{LR}(t)}{h},
\end{eqnarray}
and the diffusion constant is defined by Einstein's relation
\begin{eqnarray}
  \mathcal{D}=\frac{\sigma}{\chi},
\end{eqnarray}
where $ \chi $ is the static susceptibility.
Finally, in the case of ballistic transport the Drude weight corresponds to
\begin{eqnarray}
  D=\lim_{t\to\infty} \lim_{n\to\infty}\lim_{h\to 0} \frac{j_{LR}(t)}{t\, h}.
\end{eqnarray}

\section{The~spatio-temporal correlation function}\label{app:structureFactor}
Here we present some of the~details of calculation that were omitted
in~section~\ref{sec:StructureFactor}.
All the~nonvanishing submatrices~$\mathcal{U}^{d\,d^{\prime}}$~\eqref{eq:structureFactorReducedPropagator}
are block 3-diagonal.
The~blocks of the submatrix ~$\mathcal{U}^{00}$ are of the size~$2\times 2$ and
the~blocks in~$\mathcal{U}^{d\,0}$ for~$d\neq0$ have the dimension~$2\times 1$,
\begin{eqnarray}\fl \label{eq:zeroblocks}
  \eqalign{
    \eqalign{
      \mathcal{U}^{00}&=
      L\otimes
      \begin{bmatrix}
        (1-\rho)^2 & \rho(1-\rho)\\
        0&0\\
      \end{bmatrix}+
      I\otimes
      \begin{bmatrix}
        \rho^2 & \rho(1-\rho)\\
        \rho(1-\rho) & \rho^2\\
      \end{bmatrix}+\\
      &+ U\otimes
      \begin{bmatrix}
        0&0\\
        \rho(1-\rho)&(1-\rho)^2\\
      \end{bmatrix},
    }\\
    \mathcal{U}^{10} = I\otimes
    \begin{bmatrix}
      -(1-\rho)^2& -\rho(1-\rho)\\
    \end{bmatrix}+
    U\otimes
    \begin{bmatrix}
      \rho(1-\rho) & (1-\rho)^2\\
    \end{bmatrix},\\
    \mathcal{U}^{-1 0} = L\otimes
    \begin{bmatrix}
      (1-\rho)^2 & \rho(1-\rho)\\
    \end{bmatrix}+I\otimes
    \begin{bmatrix}
      -\rho(1-\rho) & -(1-\rho)^2\\
    \end{bmatrix},\\
    \mathcal{U}^{20} = I\otimes
    \begin{bmatrix}
      -\rho (1-\rho) & \rho(1-\rho)
    \end{bmatrix},\qquad
    \mathcal{U}^{-20} = I\otimes
    \begin{bmatrix}
      \rho (1-\rho) & -\rho(1-\rho)
    \end{bmatrix},\\
    \mathcal{U}^{30} = U\otimes
    \begin{bmatrix}
      -(1-\rho)^2 & (1-\rho)^2\\
    \end{bmatrix},\qquad
    \mathcal{U}^{-30} = L\otimes
    \begin{bmatrix}
      (1-\rho)^2 & -(1-\rho)^2\\
    \end{bmatrix},
  }
\end{eqnarray}
where we defined
\begin{eqnarray}\fl
  L=\begin{bmatrix}
    \ddots & \\
    1 & 0 & \\
    & 1 & 0 & \\
    & & 1 & 0 & \\
    & & & & \ddots\\
  \end{bmatrix},\quad
  I=\begin{bmatrix}
    \ddots & \\
    & 1 & \\
    & & 1 & \\
    & & & 1 & \\
    & & & & \ddots\\
  \end{bmatrix},\quad
  U=\begin{bmatrix}
    \ddots & \\
    & 0 & 1 \\
    & & 0 & 1 \\
    & & & 0 & 1\\
    & & & & \ddots\\
  \end{bmatrix}.
\end{eqnarray}
The~other matrices~$\mathcal{U}^{d^{\prime}\,d}$ with~$d^{\prime},d\neq0$ are simpler,
\begin{eqnarray}\label{eq:nonzeroblocks}
  \begin{aligned}
    \mathcal{U}^{2d-1\,2d-2}&=\rho(1-\rho)\,L,\qquad& \mathcal{U}^{2d\,2d-2}&=\rho^2\, I,\\
    \mathcal{U}^{2d+1\,2d-2}&=(1-\rho)^2\,I,\qquad&\mathcal{U}^{2d+2\,2d-2}&=(1-\rho)^2\, U,\\
    \mathcal{U}^{2d-1\,2d-1}&=(1-\rho)^2\,L,\qquad&\mathcal{U}^{2d\,2d-1}&=\rho(1-\rho)\, I,\\
    \mathcal{U}^{2d+1\,2d-1}&=\rho^2\,I,\qquad&\mathcal{U}^{2d+2\,2d-1}&=\rho(1-\rho)\, U.
  \end{aligned}
\end{eqnarray}
Each submatrix~$\mathcal{U}^{d^{\prime}\, d}$ can be (block) diagonalized in a~similar way
as in section~\ref{sec:inhomogeneousQuench}. In the Fourier basis
\begin{eqnarray}
  \eqalign{
    e(k)=\sum_x \mathrm{e}^{\ii k x}a_x^{\text{e}\,0},\qquad
    o(k)=\sum_x \mathrm{e}^{\ii k x}a_x^{\text{o}\,0},\\
    g^{[d]}(k)=\sum_x \mathrm{e}^{\ii k x} a_x^{d},
  }
\end{eqnarray}
the~infinitely dimensional
matrices from~\eqref{eq:zeroblocks},~\eqref{eq:nonzeroblocks}, which are of the form
\begin{eqnarray}
  \mathcal{U}^{d^{\prime} d}=L\otimes m_L^{d^{\prime} d} + I\otimes m_I^{d^\prime\, d}
  + U\otimes m_U^{d^{\prime}\, d},
\end{eqnarray}
read
\begin{eqnarray}
  \widetilde{\mathcal{U}}^{d^{\prime} d}= \mathrm{e}^{-\ii k} m_L^{d^{\prime}\,d} + m_I^{d^{\prime}\,d}
  + \mathrm{e}^{\ii k} m_D^{d^{\prime}\,d}.
\end{eqnarray}
Therefore the~reduced propagator~$\mathcal{U}$ takes the form~\eqref{eq:InfFourierMatrixBlocks}.
\end{document}